\documentstyle{article}
\input FEYNMAN
\begin{document}
\title{
\begin{flushright}
\normalsize
GUTPA/97/05/4 \\
NBI--HE--97--24 \\
hep--ph/9706212
\end{flushright}
\vspace{0.5cm}
Fermion masses, neutrino mixing and CP violation 
from the anti-grand unification model}
\author{C.D. Froggatt$^{\rm a}$\thanks{c.froggatt@physics.gla.ac.uk} \and
M. Gibson$^{\rm a}$\thanks{m.gibson@physics.gla.ac.uk} \and
H.B. Nielsen$^{\rm b}$\thanks{hbech@nbi.dk} \and
D.J. Smith$^{\rm b}$\thanks{D.Smith@nbi.dk}}
\maketitle
{\flushleft $^{\rm a}$ Department of Physics and Astronomy,
University of Glasgow, Glasgow, G12 8QQ, UK}
\vspace{0.2cm}
{\flushleft $^{\rm b}$ Niels Bohr Institute, Blegdamsvej 17-21,
DK 2100 Copenhagen, Denmark}
\vspace{0.5cm}
\begin{abstract}
The fermion masses and mixing angles are fitted 
using only 3 free parameters in a
non-supersymmetric extension of the Standard Model, with
new approximately conserved chiral gauge quantum numbers
broken by a set of Higgs fields. 
The fundamental mass scale of this anti-grand unification model
is given by the Planck mass. We also calculate
neutrino mixing angles and masses, as well 
as CP violation from the CKM matrix.
A good fit is obtained to the observed fermion 
masses but our predictions of the
neutrino masses are too small to lead to any 
observable neutrino oscillation effects claimed today,
without introducing another mass scale. 
We also give some arguments in support of this type of model
based on the observed fermion masses.
\end{abstract}

\section{Introduction}

In a previous paper \cite{SMG3U1} we presented a model to explain 
the origin of the Standard Model (SM) fermion mass and mixing 
hierarchy, based on the so-called anti-grand unified extension
of the SM gauge group (SMG). This model gives rise to a characteristic 
structure for the quark-lepton mass matrices; in this paper we shall 
try to argue that some features of this model are implied by 
the experimentally determined fermion masses. We shall also extend 
the analysis in \cite{SMG3U1} to include CP violation due to a 
complex phase in the CKM quark mixing matrix and to calculate the 
neutrino mass matrix to get predictions for neutrino masses and 
mixing angles.

In section~\ref{model} we will describe how to calculate fermion 
masses and mixing angles when viewing the SM as an effective theory. 
The observed fermion masses and mixing angles will be discussed 
in section~\ref{expmass}. An approximate parameterisation of the 
masses will then be described. In section~\ref{boot} we will give 
a simple method by which a model could naturally explain such a 
parameterisation. Our anti-grand unified model will be described
in section~\ref{AGUT}, where we show that it fulfils the 
requirements of section~\ref{boot}. The gauge quantum numbers 
of the quarks and leptons in our model
are fixed by the requirement of anomaly cancellation. 

In section~\ref{construct} we will 
describe the construction of our model for the second and third 
generation fermion masses and mixing angle, 
in terms of two Higgs fields 
in addition to the usual Weinberg-Salam (WS) Higgs field.
We emphasise in section~\ref{predictions} the general features, 
suggested by phenomenology and the anti-grand unified model, 
which underlie the structure of our mass matrices for the 
second and third generation fermions. 
The choice of the Higgs field quantum numbers 
and the extension of the mass matrices to include the first 
generation particles are discussed in section~\ref{higgs}.  
This leads to definite predictions for the order of magnitude 
of the fermion masses and mixing angles in terms of the vacuum 
expectation values (VEVs) of the Higgs fields. In 
section~\ref{results} we present our best fit to the 
conventional experimental masses and mixing angles, 
in terms of just three VEVs of the same order of magnitude. 
We also make a fit using preliminary values 
for the light quark masses 
extracted from lattice QCD. These smaller ``lattice'' values 
for the light quark masses motivated us to consider, in 
section~\ref{alternative}, the possibility of an alternative 
mass matrix structure

Then we discuss other predictions of our model. In section~\ref{CP} 
we consider how much CP violation would be expected from the 
CKM matrix in our model. Finally we turn to the neutrino masses 
and mixing angles in section~\ref{neutrinos}. The effective Majorana 
mass matrices for the three light neutrinos in models with a 
hierarchical mass matrix can naturally have quasi-degenerate 
mass eigenstates with maximal mixing \cite{fn2}. Maximal 
neutrino mixing provides a candidate explanation for the 
atmospheric muon neutrino deficit or the solar neutrino problem 
with vacuum oscillations. 
Our model does have such a structure but, 
with a see-saw mass scale equal to the Planck scale, the predicted 
neutrino masses are too small to give observable effects. As 
in most other models, it is necessary to introduce an intermediate 
mass scale in order to obtain observable neutrino mixing. This is 
not so attractive in the anti-grand unification model. However 
we can obtain identical structure for the fermion mass matrices 
in an anomaly free ${\rm SMG} \otimes {\rm U}(1)^3$ model, 
in which the 
fundamental scale is unconstrained and could be taken as 
intermediate between the electroweak and Planck scales.
We present our conclusions in section~\ref{conclusions}.

\section{Modelling fermion masses}
\label{model}

In the SM the fermions (apart from the neutrinos) 
all get a mass via the Higgs
mechanism. Before electroweak symmetry breaking 
there are interactions of the
form:
\begin{equation}
{\cal L}_{{\rm mass}} = \overline{Q_L}H_U\tilde\Phi_{WS}U_R +
	\overline{Q_L}H_D\Phi_{WS}D_R +
	\overline{L_L}H_E\Phi_{WS}E_R + {\rm h.c.}
\label{L_Higgs}
\end{equation}
where $\Phi_{WS}$ is the WS Higgs field, 
$Q_L$ is the 3 SU(2) doublets of
left-handed quarks, $H_U$ is the $3 \times 3$ 
Yukawa matrix for the up-type
quarks, etc. If we represent the SU(2) 
doublets $\Phi_{WS}$ and $Q_L$ as 2
component column vectors, we define:
\begin{equation}
\tilde\Phi_{WS} = \left ( \begin{array}{cc} 0 & 1 \\ 1 & 0 \end{array}
			 \right ) \Phi_{WS}^{\dagger}
\end{equation}
and
\begin{equation}
\overline{Q_L} =
	\overline{ \left ( \begin{array}{c} U_L \\ 
D_L \end{array} \right ) }
	= ( \overline{U_L} \; \overline{D_L} )
\end{equation}
where $\overline{U_L}$ are the CP conjugates of the 
3 left-handed up-type quarks.

After electroweak symmetry breaking the WS Higgs field gets a
vacuum expectation value (VEV) and we would write:
\begin{equation}
{\cal L}_{{\rm mass}} = \overline{U_L}M_UU_R +
        \overline{D_L}M_DD_R +
        \overline{E_L}M_EE_R + {\rm h.c.}
\label{L_Mass}
\end{equation}
where the mass matrices are related to the Yukawa matrices 
and WS Higgs VEV by:
\begin{equation}
M = H \frac{\langle \phi_{WS}\rangle}{\sqrt{2}}
\label{mass-scale}
\end{equation}
and we have chosen the normalisation so that:
\begin{equation}
\langle \phi_{WS}\rangle =  246 \; {\rm GeV}
\label{WS-vev}
\end{equation}

The masses of e.g.\ the 3 up-type fermions are 
obtained from $M_U$ by
diagonalising the matrix to find the 3 eigenvalues. 
In particular we can find
unitary matrices $V_U$ and $V_D$ so that:
\begin{eqnarray}
V_U^{\dagger}M_UM_U^{\dagger}V_U & = & 
\mbox{diag}\{m_u^2,m_c^2,m_t^2\} \\
V_D^{\dagger}M_DM_D^{\dagger}V_D & = & 
\mbox{diag}\{m_d^2,m_s^2,m_b^2\}
\end{eqnarray}
The quark mixing matrix is then defined as:
\begin{equation}
V_{{\rm CKM}} = V_U^{\dagger}V_D
\end{equation}

This is where we find
aesthetic problems with the SM. First, we have 
no way of calculating the mass
matrices since there are no constraints due to 
e.g.\ gauge symmetries. The
elements of the 3 Yukawa matrices are allowed to be 
arbitrary complex numbers. This means
that there is no understanding of the origin of the 
masses or mixing angles
within the SM. The second problem is that we would 
expect that, since there is
no distinction between the 3 generations, 
the order of magnitude of the mass of
each fermion would not depend on which 
generation it was in. Further, since,
as far as the Higgs sector is concerned, 
there is no distinction between the
different types of fermions, we would 
generally expect that all fermions in
the SM (except for the neutrinos) would 
have the same order of magnitude
masses. Indeed naturality would suggest 
that all the Yukawa matrix
elements be of order unity and the fermion masses of order
$\langle \phi_{WS}\rangle/\sqrt{2} = 174$ GeV. 
This is clearly not the case.

So in order to understand the masses of the 
fermions we must postulate some
model beyond the SM, which can give different 
SM fermions very different masses
without the need for arbitrarily small 
Yukawa couplings. One method to do this
is to extend the SM gauge group so that 
the 3 generations are not equivalent
under the more fundamental gauge group. Then we would consider
eq.~(\ref{L_Higgs}) to be the effective Lagrangian 
for the Yukawa-Higgs sector. In the
full gauge group such terms would not generally 
be gauge invariant since the
SM fermions would have extra ``charges'' 
(generally Abelian and non-Abelian).
Of course such terms would still be expected in 
the effective theory at
energies much below the symmetry breaking
scale of the full gauge group to the SMG. 
The difference is that
now we have no reason to expect that all 
the Yukawa interactions in the effective theory
should be of the same order of magnitude.

\begin{figure}
\begin{picture}(40000,13000)
\THICKLINES

\drawline\fermion[\E\REG](5000,1500)[7000]
\drawarrow[\E\ATBASE](\pmidx,\pmidy)
\global\advance \pmidy by -2000
\put(\pmidx,\pmidy){$b_L$}

\put(12000,0){$\lambda_1$}

\drawline\fermion[\E\REG](12000,1500)[7000]
\drawarrow[\E\ATBASE](\pmidx,\pmidy)
\global\advance \pmidy by -2000
\put(\pmidx,\pmidy){$M_F$}

\put(19000,0){$\lambda_2$}

\drawline\fermion[\E\REG](19000,1500)[7000]
\drawarrow[\E\ATBASE](\pmidx,\pmidy)
\global\advance \pmidy by -2000
\put(\pmidx,\pmidy){$M_F$}

\put(26000,0){$\lambda_3$}

\drawline\fermion[\E\REG](26000,1500)[7000]
\drawarrow[\E\ATBASE](\pmidx,\pmidy)
\global\advance \pmidy by -2000
\put(\pmidx,\pmidy){$b_R$}

\drawline\scalar[\N\REG](12000,1500)[5]
\global\advance \pmidx by 1500
\global\advance \pmidy by 1500
\put(\pmidx,\pmidy){$\Phi_{WS}$}
\global\advance \scalarbackx by -530
\global\advance \scalarbacky by -530
\drawline\fermion[\NE\REG](\scalarbackx,\scalarbacky)[1500]
\global\advance \scalarbacky by 1060
\drawline\fermion[\SE\REG](\scalarbackx,\scalarbacky)[1500]

\drawline\scalar[\N\REG](19000,1500)[5]
\global\advance \pmidx by 1500
\global\advance \pmidy by 1500
\put(\pmidx,\pmidy){$W$}
\global\advance \scalarbackx by -530
\global\advance \scalarbacky by -530
\drawline\fermion[\NE\REG](\scalarbackx,\scalarbacky)[1500]
\global\advance \scalarbacky by 1060
\drawline\fermion[\SE\REG](\scalarbackx,\scalarbacky)[1500]

\drawline\scalar[\N\REG](26000,1500)[5]
\global\advance \pmidx by 1500
\global\advance \pmidy by 1500
\put(\pmidx,\pmidy){$T$}
\global\advance \scalarbackx by -530
\global\advance \scalarbacky by -530
\drawline\fermion[\NE\REG](\scalarbackx,\scalarbacky)[1500]
\global\advance \scalarbacky by 1060
\drawline\fermion[\SE\REG](\scalarbackx,\scalarbacky)[1500]

\end{picture}
\caption{Feynman diagram for bottom quark mass in the full theory.
The crosses indicate the couplings of 
the Higgs fields to the vacuum.}
\label{MbFull}
\end{figure}
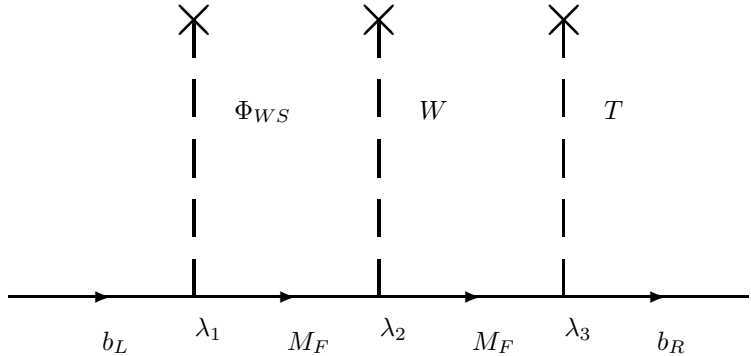

As an example consider the bottom quark. 
If we suppose that the dominant
transition between the left- and 
right-handed components in the fundamental
theory involves not only the WS Higgs field, 
but e.g.\ 2 other Higgs fields $W$ and
$T$ as shown in fig.~\ref{MbFull}, 
we would get the following relation 
\cite{fn1} for
the effective Yukawa coupling of the bottom quark in the SM:
\begin{equation}
h_b \simeq \frac{\langle W\rangle}{M_{F}}
\frac{\langle T\rangle}{M_{F}}
\label{byukawa}
\end{equation}
where $\langle W\rangle$ and $\langle T\rangle$ 
are simply the VEVs of the new
Higgs fields $W$ and $T$, and
$M_F$ is the (fundamental) mass scale 
of the intermediate fermions. Here we are
assuming that all fundamental Yukawa 
couplings (the $\lambda_i$) are of order
1. Thus the order of magnitude of the 
effective SM Yukawa coupling 
constants are given by the product of 
small symmetry breaking factors,
like $\frac{\langle W\rangle}{M_{F}}$ 
and $\frac{\langle T\rangle}{M_{F}}$.
So now we can explain why
the fermions in the SM have different masses 
and we can construct explicit
models, by choosing extended gauge quantum 
numbers for the SM fermions and
making specific choices of Higgs fields 
$W$, $T$, etc.\, to allow the
transitions in the fundamental theory 
which give masses to the SM fermions.

It is important to note that this type of 
model cannot give exact predictions,
since we are still unable to 
calculate fundamental Yukawa couplings.
Essentially we introduce a lot more 
fundamental Yukawa couplings but then make
the naturality assumption that they 
are all of order 1. However, the problem we
are addressing is the huge range of 
fermion masses and so we can get some
understanding of this hierarchy without any knowledge of 
why the masses are exactly what
they are. Furthermore we assume that 
there exists a spectrum of vector-like
fermion states, all having a mass of order $M_F$, 
which can mediate all of the
required symmetry breaking transitions.

Clearly there are many different models 
we could propose to model the fermion
masses. In order to make progress it 
is therefore necessary to examine the
experimentally measured masses and 
look for relations such as order of
magnitude degeneracy. One point to 
note is that the effective Yukawa couplings
we predict will be the values at the 
fundamental scale $M_F$ (assumed to be of
order the Planck scale $M_{Planck} \simeq 10^{19} \; {\rm GeV}$) 
rather than experimental scales such as 1 GeV.

\section{Measured fermion masses}
\label{expmass}

The masses of the fermions in the SM are 
usually quoted as running masses, 
at some scale such as 1 GeV,
except for the top mass which is generally 
quoted as a pole mass. These masses
can be evolved to other scales by using 
the renormalisation group equations
(RGEs). One of the assumptions we make 
is that the SM is valid up to the
Planck scale, but assuming a lower fundamental scale (say of order
$10^{15}$ GeV) makes no essential
difference. So we evolve the masses 
to the Planck scale to compare them. The
main effect is to change the ratios of 
the quark to lepton masses and also the
ratio of the top quark to other quark masses. 
At a scale $\mu$ we can express the
running masses in terms of the Yukawa couplings by:
\begin{equation}
m(\mu) = h(\mu)\frac{\langle \phi_{WS}\rangle}{\sqrt{2}}
\end{equation}
The pole mass of a quark is given to leading order 
in the strong coupling by:
\begin{equation}
M = m(M)\left(1+\frac{4}{3}\frac{\alpha_S(M)}{\pi}\right)
\end{equation}
In table~\ref{ExpMasses} we show typically quoted values for 
the SM fermion Yukawa couplings at 1 GeV
\cite{PDG} and the
corresponding Yukawa couplings at the Planck scale 
evolved using the 1-loop
RGEs for the SM (see e.g.\ \cite{RGEs}).
There is some ambiguity, particularly for the light quarks, in 
extracting these Yukawa couplings from experiment. We shall consider
alternative smaller values for the $u$, $d$ and $s$ quark masses,
suggested by recent lattice calculations \cite{udsMasses}, 
in section~\ref{results}.  

\begin{table}
\caption{Yukawa couplings and running masses 
of the SM fermions at 1 GeV and
Yukawa couplings at the Planck scale. The top quark pole mass is 
taken to be 180 GeV.}
\begin{displaymath}
\begin{array}{ccccc} \hline
{\rm Fermion} & m(1 {\rm \; GeV}) & h(1 {\rm \; GeV}) & 
h(M_{Planck}) & \ln(h(M_{Planck})) \\ \hline
u & 4 {\rm \; MeV} & 2.3 \times 10^{-5}	& 
5.4 \times 10^{-6} & -12.1 \\
c & 1.4 {\rm \; GeV} & 8.0 \times 10^{-3} & 
1.9 \times 10^{-3} & -6.3 \\
t & 240 {\rm \; GeV} & 1.38 & 0.39 & -0.9 \\
d & 9 {\rm \; MeV} & 5.2 \times 10^{-5}	& 
1.3 \times 10^{-5} & -11.3 \\
s & 200 {\rm \; MeV} & 1.1 \times 10^{-3} & 
2.8 \times 10^{-4} & -8.2 \\
b & 6.3 {\rm \; GeV} & 3.6 \times 10^{-2} & 
7.4 \times 10^{-3} & -4.9 \\
e & 0.5 {\rm \; MeV} & 2.9 \times 10^{-6} & 
2.9 \times 10^{-6} & -12.8 \\
\mu & 105 {\rm \; MeV} & 6.0 \times 10^{-4} & 
5.9 \times 10^{-4} & -7.4 \\
\tau & 1.78 {\rm \; GeV} & 1.0 \times 10^{-2} & 
1.0 \times 10^{-2} & -4.6 \\
	\hline
\end{array}
\end{displaymath}
\label{ExpMasses}
\end{table}

In table~\ref{ExpMixing} we show the 
magnitudes of the 3 mixing angles
(without the CP phase) at 1 GeV and the Planck scale. 
We can see that $V_{us}$
does not change significantly, 
but the other 2 mixing angles are different at
the two scales. We find that the running is 
independent of the CP phase chosen.
We also find, in agreement with \cite{RunCKM}, 
that the ratio of $V_{cb}$ to
$V_{ub}$ is approximately constant. 
The difference at the two scales is only
sensitive to the top quark mass.
\begin{table}
\caption{Experimental values of 
the mixing angles at 1 GeV and the Planck
scale.}
\begin{displaymath}
\begin{array}{cccc}
\hline
{\rm Mixing \; angle} & V(1 \; {\rm GeV}) & V(M_{Planck}) 
& ln(V(M_{Planck})) \\
\hline
V_{us} & 0.22 & 0.22 & -1.5 \\
V_{cb} & 0.041 & 0.049 & -3.0 \\
V_{ub} & 0.0035 & 0.0042 & -5.5 \\
\hline
\end{array}
\end{displaymath}
\label{ExpMixing}
\end{table}

If we now look for order of magnitude equalities at the Planck scale 
we can see that, with the exception of the top 
and probably also the charm quarks, the
Yukawa couplings of the fermions within each generation are order of
magnitudewise degenerate. 
We could explain the down-type quarks and the leptons
having degenerate masses in a grand-unified model 
such as the well known SU(5)
model. However, in reality, the degeneracy 
is only true in an order of
magnitude approximation, whereas we would 
expect exact equality at some scale
in such a model\footnote{It is of course possible to 
avoid the predicted degeneracy by introducing a non-minimal 
Higgs structure\cite{gj}.} 
Another feature, which indicates that the degeneracies are not
due to unification, is that the up quark is 
approximately degenerate with the down quark and the 
electron but the charm and, most obviously, the top 
quarks do not fall into the
same pattern. Obviously we could
say that the up quark is the exception 
and it is simply chance that it has the
same order of magnitude mass as the down quark 
and the electron, but this is not entirely
satisfactory. For example, if the up-type 
quarks get masses by a different
mechanism (e.g.\ supersymmetry with a large $\tan \beta$ 
which could produce the
observed top and bottom quark masses 
even if they actually had similar Yukawa
couplings), then it would seem to be 
natural to expect that the up quark would
be heavier than the down quark and the 
electron, just as the charm and top quarks are
heavier than the other fermions in their generations. 
Therefore we take the view that there is an 
approximate mass degeneracy between charged fermions 
within each generation, which is certainly true
for the lightest generation. However, 
in the other two generations, the
charm and top quarks are exceptions to this rule. 
This means that the charm
and top quarks must get their masses by a different 
mechanism from the other fermions.

\section{A Natural Explanation}
\label{boot}

If we consider the hypothetical situation where the charged 
fermions within each
generation have the same masses, then 
it would be natural to conclude that the
masses were generated by some mechanism 
which didn't distinguish between each
type of fermion, but did distinguish 
between fermions in different generations. 
This would indicate that, 
in some model more fundamental than the SM, there
should be a distinction between the 3 generations. 
We shall refer to these
distinct ``generations'' in the fundamental 
theory as ``proto-generations''.
We cannot really
claim that the SM fermion masses could be 
described by an approximation to such
a model, since the charm and top quarks 
are clearly not even order of magnitudewise
degenerate with the other fermions in their 
generations. However, we shall show
that the SM fermion masses can be described 
by a very similar type of model.

An important point to consider is that 
the fermion masses actually tell us
only what the eigenvalues of the mass 
matrices are, they do not specify the
complete matrices. It seems reasonable 
to assume that in a model where each
generation is distinguished (e.g.\ by 
having different charges under some new
gauge interaction), the dominant elements 
in the mass matrices which would
lead to the fermion masses could easily 
be elements on the diagonal, i.e.\
the transitions between left- and 
right-handed fermions within the same
generation. Now we could explain 
approximate degeneracy of fermions within
each generation by a model which produced 
mass matrices with the same order
of magnitude diagonal elements. 
The important point is that we need not
require the complete matrices to 
be similar, as long as the off-diagonal
elements make no significant contributions to the eigenvalues.

Now we can explain why we have the 
exceptions of the top and charm
quarks. To first approximation the 
largest eigenvalue of a matrix is simply
the largest element. So in the down-type 
and charged lepton mass matrices,
$M_D$ and $M_E$, the
largest element is the one corresponding to 
the interaction between the left-
and right-handed fermions in the 3rd 
proto-generation. In the up-type mass
matrix, $M_U$, it happens that our assumption about 
diagonal elements being dominant
was wrong. This is not totally surprising, 
since we don't really have any good
reason to make such an assumption 
without knowledge of the fundamental
mass-generating mechanism. Since we know 
that the left-handed top
and bottom quarks are in the same SU(2) 
doublet in the SM, this means that
the right-handed top quark is not really the fermion in the 3rd
proto-generation of this model. The 
right-handed top quark must therefore
belong to the 1st or 2nd proto-generation. 
This makes no difference in the
SM since there is no distinction between 
the generations, but in such extended
models a distinction does exist. By 
identifying the right-handed top quark
with the 2nd proto-generation and 
the right-handed charm quark with the 3rd
proto-generation, we have a simple mechanism 
to explain why their masses differ
from the other fermions within their generations.

So now we have some important criteria for a 
model of the fermion masses.
We have made the assumption that the SM is 
valid up to some high scale, such
as the Planck scale. We have shown 
that there are several order of magnitude
degeneracies when the SM fermion Yukawa 
couplings are compared at this scale.
This has led to the requirement that 
the generations should be distinguished
in the fundamental model, and that 
the fermions within each of these
proto-generations should have order of magnitude degenerate masses.
However, the interactions between generations 
must distinguish between the
charged leptons and the 2 different types of quarks, 
so that the top and
charm quark masses can be explained as 
originating from interactions between left-
and right-handed fermions from different 
proto-generations (most simply being
due to the fact that the right-handed 
top quark of the SM actually belongs to
the 2nd proto-generation and the right-handed charm quark to the 3rd
proto-generation). Note that this mechanism would not work if,
for example, only the
top quark mass deviated from the ``rule'' of degeneracy within generations.

\section{The anti-grand unification model}
\label{AGUT}
We will now propose a model which has all 
the features proposed in the
previous section. We will first describe 
the model and then show that it
does in fact satisfy all our requirements. 
The model we are considering is
called the anti-grand unification (AGUT) model. 
It has previously been
considered as a candidate for explaining the 
fermion masses \cite{Gerry2}.
The gauge group for the model is:
\begin{equation}
{\rm G}={\rm SMG}_1 \otimes {\rm SMG}_2 \otimes 
{\rm SMG}_3 \otimes {\rm U}(1)_f
\end{equation}
where we have defined:
\begin{equation}
{\rm SMG}_i={\rm SU}(3)_i \otimes {\rm SU}(2)_i \otimes
		{\rm U}(1)_i
\end{equation}
The three ${\rm SMG}_i$ groups will be 
broken down to their diagonal
subgroup, which is the gauge group of the SM. 
The ${\rm U}(1)_f$ group will
be totally broken. So rather than unifying 
the simple factors of the SMG, the
AGUT model ``splits'' them into 3 copies 
of each (hence the name of the model).
This gauge group (strictly speaking 
without the ${\rm U}(1)_f$ group) has
been used to successfully predict the 
values of the running gauge coupling
constants in the SM at the Planck 
scale \cite{Bennett}, as critical couplings
estimated using lattice gauge theory.

We put the SM fermions into this group G 
in an obvious way. We have one
generation of fermions coupling to 
each ${\rm SMG}_i$ in exactly the same
way as they would couple to the SMG in 
the SM. The broken chiral gauge
quantum numbers of the quarks and 
leptons under the symmetry groups
${\rm SMG}_i$ can readily explain 
the mass differences between fermion
generations but cannot explain all 
the mass splittings within each generation,
such as the ratio of the top and bottom quark 
masses. It is for this
reason that the Abelian flavour group 
${\rm U}(1)_f$ is introduced.
We then choose ${\rm U}(1)_f$ charges 
with the constraint that there should
be no anomalies and no new mass-protected fermions.
This leads us almost uniquely to
the set of charges $y_f$ shown in table~\ref{Q_f}. 
We have labelled the fermions
coupling to ${\rm SMG}_i$ by the names of 
the `i'th generation of SM fermions.
However, this is just a method of labelling 
the representations of the full
gauge group and, as we have discussed 
in section~\ref{boot}, we expect that,
for example, the fermion we have labelled
$c_R$ will in fact turn out to be the 
right-handed top quark in the SM.

\begin{table}
\caption{${\rm U}(1)_f$ charges of the SM fermions.}
\begin{displaymath}
\begin{array}{cc|cc|cc}
\hline
{\rm Fermion} & y_f & {\rm Fermion} & y_f & 
{\rm Fermion} & y_f \\ \hline
u_L & 0 & c_L & 0 & t_L & 0 \\
u_R & 0 & c_R & 1 & t_R & -1 \\
d_R & 0 & s_R & -1 & b_R & 1 \\
e_L & 0 & \mu_L & 0 & \tau_L & 0 \\
e_R & 0 & \mu_R & -1 & \tau_R & 1 \\ \hline
\end{array}
\end{displaymath}
\label{Q_f}
\end{table}

Although the quantum numbers of the fermion 
fields are determined by the
theoretical structure of the model 
(in particular the requirement of
anomaly cancellation), we do have some freedom in the choice
of the quantum numbers of the Higgs fields. 
So, to model the fermion masses
in the SM, we must choose how to break G down to SMG.

A crucial simplification can be made 
by considering only the Abelian charges,
when formulating a model for symmetry 
breaking. We can justify this by noting
that all fermions are in singlet or 
fundamental representations of the
non-Abelian groups. In the SM the fermions 
obey a charge quantisation rule:
\begin{equation}
\frac{y}{2} + \frac{1}{2}{\rm ``duality"} +
	\frac{1}{3}{\rm ``triality"} \equiv 0 \pmod 1
\label{SMGiChQu}
\end{equation}
where $y$ is the conventional weak hypercharge. 
Here ``duality'' is 1 for the
fundamental representation of SU(2) 
(the doublet) and 0 for the singlet.
Similarly ``triality'' is 1 for the 
SU(3) triplet, -1 for the anti-triplet and
0 for the singlet. Therefore we see that, 
if these are the only allowed
non-Abelian representations, we can 
calculate the non-Abelian representations
from the weak hypercharge. Similarly in the 
AGUT model we have such a charge
quantisation for the weak hypercharge $y_i$ 
associated with each ${\rm SMG}_i$ 
separately, so that the 4 Abelian charges completely
specify the representation under the full AGUT gauge group.
We assume that the Higgs scalar fields also satisfy the 
charge quantisation rule for each $SMG_i$ and belong to singlet 
or fundamental representations of the non-Abelian groups.

As we discussed in section~\ref{boot}, we 
wish to have a model where the
diagonal elements in the different  
mass matrices $M_U$, $M_D$ and $M_E$ are the same, but off-diagonal
elements should, in general, differ 
in different matrices. We can now show that
this will always be the case in the AGUT model.

Let us define the U(1) charge vector:
\begin{equation}
\vec{Q} \equiv \left ( \frac{y_1}{2},\frac{y_2}{2},
\frac{y_3}{2},y_f \right )
\end{equation}
Then the net charges of the Higgs fields, 
other than the WS Higgs field, in a transition
from one left-handed fermion $f_L$ to a 
right-handed fermion $g_R$ of the same type, are
given by:
\begin{equation}
\Delta\vec{Q}_{fg} = \vec{Q}_{f_L} - \vec{Q}_{g_R} \pm \vec{Q}_{WS}
\label{DeltaQ}
\end{equation}
where we have a plus sign for up-type quarks 
and a minus sign for down-type
quarks and charged leptons. 
Mass matrix elements with the same value of $\Delta\vec{Q}_{fg}$ will
be mediated by the same set of Higgs fields and suppressed by 
the same product of symmetry breaking parameters.
To compare elements in the same place in different
mass matrices we will define:
\begin{equation}
\Delta\vec{Q}_{Tij} = \Delta\vec{Q}_{T_iT_j}
\end{equation}
and to consider the diagonal elements we will 
simplify the notation further by
defining:
\begin{equation}
\Delta\vec{Q}_{Ti} = \Delta\vec{Q}_{Tii}
\label{QDiag}
\end{equation}
where $T_i$ refers to the fermion in the `i'th 
proto-generation of type $T$; e.g.\
$D_2$ is equivalent to $s$, the strange quark.

Since we can use the Higgs fields $W$ and $W^{\dagger}$ 
(which have opposite charges) equivalently in
non-supersymmetric models, we will have the 
same order of magnitude diagonal
elements if:
\begin{equation}
\Delta\vec{Q}_{Ui} = \pm\Delta\vec{Q}_{Di} = \pm\Delta\vec{Q}_{Ei}
\label{DeltaQ2}
\end{equation}
for any combination of plus and minus signs 
(possibly dependent on $i$). From
eq.~(\ref{DeltaQ}) we can see that, 
if we choose both signs to be minus, the
charges of the WS Higgs field $\vec{Q}_{WS}$ 
cancel out in the relations of eq.~(\ref{DeltaQ2}). We can then easily
see from table~\ref{Q_f} that the ${\rm U}(1)_f$ charges allow:
\begin{equation}
\Delta\vec{Q}_{Ui} = -\Delta\vec{Q}_{Di} = -\Delta\vec{Q}_{Ei}
\label{QDiagEql}
\end{equation}
The charges $y_j$ for these diagonal elements 
are all 0 for $j \ne i$, so the only requirement left is that
the charges $y_i$ satisfy eq.~(\ref{QDiagEql}). 
This is true for the simple
reason that these are the usual weak hypercharges 
of the fermions in the SM
and the relation is just the condition 
that the fermions have the correct
charges to couple to the same WS Higgs field. 
Therefore we have proved that
the diagonal elements in the 3 mass matrices $M_U$, $M_D$ and $M_E$
are the same for each
matrix, no matter what Higgs fields we choose 
or how we extend the quantum
numbers of the WS Higgs field. Furthermore 
it is fairly obvious that, in general, the
off-diagonal elements will not be the 
same in the different matrices. So the
AGUT model has satisfied the requirements 
discussed in section~\ref{boot}.

\section{Constructing a model of the Higgs fields for the 2nd and 
3rd generations}
\label{construct}

There are many ways to approach the 
construction of a realistic model of the
fermion masses. In section~\ref{boot} we 
argued that there should be approximate degeneracy
of masses between fermions within each generation, 
with the exception of the top and charm quarks. 
We explained how this could be achieved 
naturally in a model where
the diagonal elements of each mass matrix 
were of the same order of magnitude
but, in general, other elements were 
different in each matrix. As we noted in
section~\ref{AGUT}, this is precisely 
the case in the AGUT model. Of course
there are other possible models 
where the mass matrices are of this form.
However, one important point about the 
AGUT model is that we have not enforced
such a condition; rather it was forced 
upon us by the requirement of anomaly
cancellation, which determined the charges 
of the fermions. So, however we
approach the construction of a definite model (by
making a definite choice of the Higgs 
fields to break the AGUT gauge group
down to the SMG), there is always the possibility of natural order 
of magnitude degeneracies between the masses of 
different fermions within the same generation. Whether or
not this actually occurs depends on 
whether or not the appropriate diagonal
elements give the dominant contribution to 
the eigenvalue of the relevant matrices.

We have previously presented one method 
for constructing a realistic model of
the fermion masses and mixing angles \cite{SMG3U1}. 
Here we will give an
alternative method which leads to the same model. 
First we shall choose the
quantum numbers of the WS Higgs field so that this is the
only Higgs field required to produce the top quark 
mass and, thereby, the top quark Yukawa
coupling is naturally of order one. As we have explained in
section~\ref{boot}, in this type of model 
the dominant element in the up-type
mass matrix cannot be a diagonal element, 
since then we could not predict that
the top quark was much heavier than 
the bottom quark and tau lepton. So we choose
the element corresponding to a transition
between $t_L$ and $c_R$, as defined in section~\ref{AGUT},
to be dominant. Correspondingly the
element which determines the charm quark mass 
corresponds to the transition between
$c_L$ and $t_R$. So all we have really 
done is relabel the right-handed top
and charm quarks which is simply a 
matter of definition (identification of
fermions in our model with those of the SM), 
since they have the same gauge
representations in the SM.

We could in fact choose the dominant element to be any of the 6
off-diagonal elements in the up-type mass matrix. 
However, interchanging
$SMG_2$ and $SMG_3$ along with a change of sign of the
${\rm U}(1)_f$ charges simply corresponds 
to relabelling the second and
third proto-generations. This relates 
the 6 elements in pairs so that there
are only really 3 distinct choices to be made. 
It turns out that, with
our assumption that the first generation 
fermions are all order of magnitudewise
degenerate in mass and that their masses 
are due to the same diagonal
element in the three mass matrices, the 
3 choices lead to equivalent models
for the second and third generation masses. 
This is because, after making the
appropriate choice of WS Higgs quantum numbers, 
the quantum numbers of the
transitions among the second and 
third generations are the same in all 3
cases up to permutations of the $SMG_i$ 
and rescaling of the ${\rm U}(1)_f$
charges. However, the quantum numbers 
in transitions involving the first
generation are dependent on which of the 
3 choices is made. These charges
cannot be related to a different case by 
such a linear transformation of the U(1) charges.
Clearly the 3 choices
correspond to the arbitrary labelling 
of the 3 proto-generations according to
the ${\rm U}(1)_f$ charges. There are 3 
rather than 6 choices since changing
the sign of the ${\rm U}(1)_f$ charges 
simply corresponds to interchanging
the 2nd and 3rd proto-generations.
So, although we will only consider the choice
made above (a dominant $t_L-c_R$ mass matrix element) 
in this paper, our conclusions about modelling the second and
third generations will be independent of this choice.

To begin with we shall concentrate on the mass matrix elements 
responsible for the second and third generation masses.
We have chosen the elements in the up-type 
mass matrix which will give the
top and charm quark masses. In 
the down-type and lepton mass matrices 
we must choose the diagonal elements, 
if we want to obtain the order of 
magnitude degeneracies between the 
bottom quark and tau lepton masses, and between the 
strange quark and muon masses. 
Therefore, ignoring the elements involving the first
generation, we have the following order of 
magnitude elements in the mass matrices:
\begin{eqnarray}
M_U & \simeq & \left ( \begin{array}{cc} m_s & m_c \\
		m_t & m_b \end{array} \right ) 
\label{Up23} \\
M_D & \simeq & \left ( \begin{array}{cc} m_s & m_bV_{23} \\
		?  & m_b \end{array} \right ) 
\label{Down23} \\
M_E & \simeq & \left ( \begin{array}{cc} m_s &  ?  \\
		?  & m_b \end{array} \right )
\label{Elec23}
\end{eqnarray}
where $m_{\mu} \simeq m_s$ and $m_{\tau} \simeq m_b$ 
and we have indicated
the likely position of the element which leads to the
mixing $V_{23}$ between second and third generation quarks.

Now we are in a position to choose specific Higgs 
fields which could lead to
these types of mass matrices. We can 
do this by noting that there are relations
between the quantum numbers of different 
elements. In particular we define:
\begin{equation}
\vec{b} = \vec{Q}_{b_L} - \vec{Q}_{b_R} - \vec{Q}_{WS}
\end{equation}
which are the total charges carried by the Higgs fields,
other than the WS Higgs field, 
in the transition which leads 
to the element in the down-type mass matrix
corresponding to the bottom quark mass. The U(1) 
charges of the WS Higgs field are:
\begin{equation}
\vec{Q}_{WS} = \vec{Q}_{c_R} - \vec{Q}_{t_L} =
	\left ( 0,\frac{2}{3},0,1 \right ) - 
	\left ( 0,0,\frac{1}{6},0 \right )
	= \left ( 0,\frac{2}{3},-\frac{1}{6},1 \right )
\label{qws}
\end{equation}
We define similar quantities for other
elements in the down-type and lepton mass 
matrices. In the up-type mass matrix
we have a different interaction with the WS Higgs 
field, and so we define the total charges carried by 
the other Higgs fields in, for example, the transition 
responsible for the charm quark mass by:
\begin{equation}
\vec{c} = \vec{Q}_{c_L} - \vec{Q}_{t_R} + \vec{Q}_{WS}
\end{equation}
Here we have used the requirement that the 
charm quark mass should be due to the
off-diagonal element corresponding to the 
transition between $c_L$ and $t_R$,
rather than the diagonal element corresponding 
to the transition between $c_L$ and $c_R$. 
It is easy to verify the following 
relations between these charge vectors:
\begin{eqnarray}
\vec{b} + \vec{c} + \vec{s} & = & \vec{0} \label{bcs}\\
\vec{Q}_{b23} + \vec{s} - \vec{b} & = & \vec{0} \label{bsV}
\end{eqnarray}
where $\vec{Q}_{b23}$ is defined to be the 
charges corresponding to the element
we expect to have magnitude $m_bV_{23}$ in eq.~(\ref{Down23}).
This means that these vectors form triangles in charge space. 
Thus we predict that (in units of the supposed 
unsuppressed top quark mass)
the mass of the bottom, charm or strange quark is no
less than the product of the other two masses. 
This is in fact true in nature
and also for the other case, eq.~(\ref{bsV}), 
with $m_b$, $m_s$ and $m_bV_{23}$.

We now try to fit the effective SM Yukawa 
couplings in terms of as few Higgs fields as possible. 
In the case of only 1 Higgs field $W$ the prediction
from eq.~(\ref{bcs}) for $h_s$, the smallest Yukawa coupling
of the three, is:
\begin{equation}
h_s \simeq h_bh_c
\label{h_s}
\end{equation}
To see this note that each Yukawa coupling is of the form
\begin{equation}
h_X \sim \langle W\rangle^{|n_X|}
\label{hWn}
\end{equation}
where from eq.~(\ref{bcs})
\begin{equation}
n_b + n_c + n_s = 0 
\label{m_bcs}
\end{equation}
and so, since $h_s<h_c<h_b$ implies $|n_s|>|n_c|>|n_b|$:
\begin{equation}
|n_s| = |n_b| + |n_c|
\label{n_s}
\end{equation}
Eq.~(\ref{hWn}) shows that eqs.~(\ref{n_s}) and 
(\ref{h_s}) are equivalent.
The VEV $\langle W\rangle$ (in units of $M_F$) 
is the suppression factor generated 
each time that the Higgs field $W$ has to be applied. 
Now eq.~(\ref{h_s}) is not true in nature (see table~\ref{ExpMasses})
and so we must use at least 2 Higgs fields. 

Whenever we have relations such as eqs.~(\ref{bcs}) and (\ref{bsV}) 
between three matrix element charge vectors,
we imagine that we can re-use chains of Higgs fields
(see fig. 1) able to exchange the 
quantum numbers necessary to give masses 
to, say, the $b$ and $c$ quarks to also give mass to the $s$ quark. 
Naturally, there will be a piece of chain common, up to charge 
conjugation, for $b$ and $c$ not used by $s$.
We seek to symbolise this chain structure in figure 2: the connected 
single lines represent logarithms of the suppression factors 
for the masses of the $b$, $c$ and $s$ quarks. The lengths 
of the double lines represent the logarithms of the bunches of 
Higgs field suppression factors 
$\frac{\langle H \rangle}{M_F}$ common for those two 
quark masses represented by the lines forming the double lines 
in question. A priori we have the possibility that, even for three 
linearly dependent transition quantum number sets (obeying, say, 
eq.~(\ref{bcs}) like $\vec{b}$, $\vec{c}$, $\vec{s}$), there 
could be some Higgs fields which are used in only one of the three 
relevant diagrams like figure 1. However this is not possible 
for a set of Higgs fields with linearly independent quantum numbers.

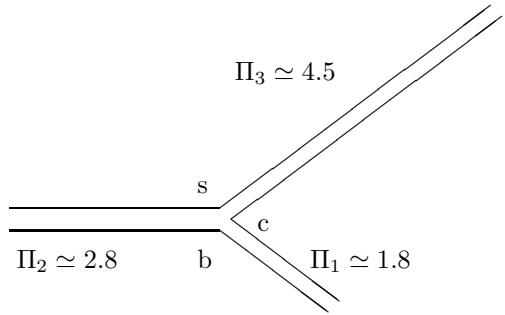
\begin{figure}

\setlength{\unitlength}{0.1mm}

\begin{picture}(1000,700)

\put(50,300){\line(1,0){280}}
\put(50,330){\line(1,0){280}}
\put(330,300){\line(4,-3){144}}
\put(345,315){\line(4,-3){144}}
\put(345,315){\line(4,3){360}}
\put(330,330){\line(4,3){360}}
\put(300,250){b}
\put(380,300){c}
\put(300,350){s}
\put(60,250){$\Pi_2 \simeq 2.8$}
\put(450,250){$\Pi_1 \simeq 1.8$}
\put(350,500){$\Pi_3 \simeq 4.5$}

\end{picture}

\caption{The magnitudes of the logarithms of the Planck scale 
Yukawa couplings for b (really $\tau$), c and s (really $\mu$) 
are represented by the total lengths of the corresponding 
single lines. The lengths of the double lines labelled by
$\Pi_{i}$ represent $-\ln \Pi_{i}$.}
\label{bcs_triangle}
\end{figure}

We now assume that just 2 Higgs fields, with linearly independent 
quantum numbers, are needed to give the masses of the $b$, $c$ and 
$s$ quarks and denote them by $W$ and $T$. Denoting the quantum 
number vectors of these fields $W$ and $T$ by $\vec{Q_W}$ and 
$\vec{Q_T}$ respectively, we must be able to write
\begin{equation}
\vec{b} = n_b\vec{Q_W} + p_b\vec{Q_T}
\end{equation}
and similar expressions for $\vec{c}$ and $\vec{s}$. Then
eq.~(\ref{m_bcs}) is satisfied as above and similarly
\begin{equation}
p_b + p_c + p_s = 0
\end{equation}
So, for each Higgs field, we must have a relation like:
\begin{equation}
|p_c| = |p_b| +|p_s|  
\end{equation}
or a similar relation with the quark names permuted. It 
follows that the suppression factors associated with each
Higgs field can be factorised in a similar way to eq.~(\ref{h_s}), 
and the 3 Yukawa couplings can be split order of magnitudewise 
in the following way:
\begin{eqnarray}
h_b & \simeq & \Pi_1\Pi_2 \label{Pi_12} \\
h_c & \simeq & \Pi_1\Pi_3 \label{Pi_13} \\
h_s & \simeq & \Pi_2\Pi_3 \label{Pi_23}
\end{eqnarray}
The $\Pi_i$s are products of Higgs suppression factors, 
i.~e.\ they are of the form:
\begin{equation}
\Pi_i = \langle W\rangle^{|n_i|}\langle T\rangle^{|p_i|}
\label{Pi_i}
\end{equation}
Here, and henceforth, we express the Higgs field VEVs 
(apart from $\langle \phi_{WS} \rangle$) in units of $M_F$.
 
Similarly eq.~(\ref{bsV}) requires the splittings:
\begin{eqnarray}
h_b & \simeq & \Pi_4\Pi_5 \label{Pi_45} \\
h_s & \simeq & \Pi_4\Pi_6 \label{Pi_46} \\
h_bV_{23} & \simeq & \Pi_5\Pi_6 \label{Pi_56}
\end{eqnarray}

\begin{figure}

\setlength{\unitlength}{0.1mm}

\begin{picture}(1000,800)

\put(50,300){\line(1,0){250}}
\put(50,330){\line(1,0){250}}
\put(300,300){\line(4,-3){168}}
\put(315,315){\line(4,-3){168}}
\put(315,315){\line(4,3){424}}
\put(300,330){\line(4,3){424}}
\put(270,250){b}
\put(370,310){s}
\put(250,370){b$V_{23}$}
\put(70,250){$\Pi_5 \simeq 2.5$}
\put(430,250){$\Pi_4 \simeq 2.1$}
\put(340,530){$\Pi_6 \simeq 5.3$}

\end{picture}

\caption{The magnitudes of the logarithms of the Planck scale 
Yukawa couplings for b, s (really $\mu$)
and b$\times V_{23}$ 
are represented by the total lengths of the corresponding 
single lines. The lengths of the double lines labelled by
$\Pi_{i}$ represent $-\ln \Pi_{i}$.}
\label{bVs_triangle}
\end{figure}
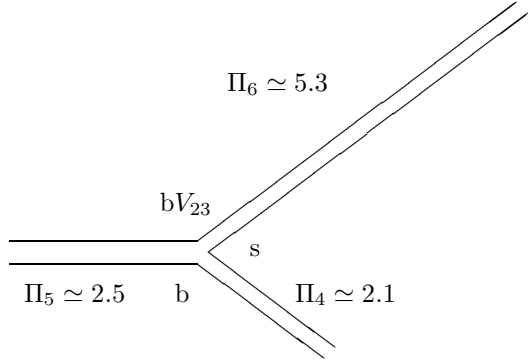

In order to motivate our final model from the numerology, we
choose to replace the quark Yukawa couplings $h_b$ and $h_s$ 
by the lepton Yukawa couplings $h_{\tau}$ and $h_{\mu}$, 
because the latter fit our model better. It is anyway our 
prediction that $h_b \simeq h_{\tau}$ and 
$h_s \simeq h_{\mu}$, since the quantum numbers satisfy: 
\begin{eqnarray}
\vec{b} & = & \vec{\tau} \\
\vec{s} & = & \vec{\mu}
\end{eqnarray}
{}From eqs.~(\ref{Pi_12})-(\ref{Pi_23}) or their illustration in 
figure 2, we see, by insertion of the Planck scale Yukawa 
couplings from table 1, that:
\begin{eqnarray}
\ln\Pi_1 & \simeq  & \frac{1}{2}(\ln h_b + \ln h_c - \ln h_s)\\
         & \simeq  & \frac{1}{2}(\ln h_{\tau} + \ln h_c - \ln h_{\mu})\\
         & \simeq & -1.8 \\
\ln\Pi_2 & \simeq  & \frac{1}{2}(\ln h_b + \ln h_s - \ln h_c)\\
         & \simeq  & \frac{1}{2}(\ln h_{\tau} + \ln h_{\mu} - \ln h_c)\\
         & \simeq & -2.8 \\
\ln\Pi_3 & \simeq  & \frac{1}{2}(\ln h_s + \ln h_c - \ln h_b)\\
         & \simeq  & \frac{1}{2}(\ln h_{\mu} + \ln h_c - \ln h_{\tau})\\
         & \simeq & -4.5
\end{eqnarray}
Similarly, from eqs.~(\ref{Pi_45})-(\ref{Pi_56}) 
or figure~\ref{bVs_triangle}, we get:
\begin{eqnarray}
\ln\Pi_4 & \simeq & -2.1 \\
\ln\Pi_5 & \simeq & -2.5 \\
\ln\Pi_6 & \simeq & -5.3
\end{eqnarray}
   
Interestingly we notice that we have a very good order 
of magnitude relation:
\begin{equation}
\Pi_3 \simeq \Pi_1\Pi_2
\end{equation}
and so we could try to consider $\Pi_1$ and $\Pi_2$ as 
single Higgs field VEVs (in units of $M_F$):
\begin{eqnarray}
\langle W\rangle & = & \Pi_1 \\
\langle T\rangle & = & \Pi_2
\end{eqnarray}
We also observe that $h_b \simeq h_{\tau}$ is, to a good numerical
approximation, split in the same way
in these 2 cases eqs.~(\ref{Pi_12}) and (\ref{Pi_45}). 
So we make the identifications:
\begin{eqnarray}
\Pi_4 & = & \Pi_1 \\
\Pi_5 & = & \Pi_2
\end{eqnarray}
This gives us the approximate values of:
\begin{eqnarray}
\langle W\rangle & \simeq & e^{-1.95} \simeq 0.14 \\
\langle T\rangle & \simeq & e^{-2.65} \simeq 0.07
\end{eqnarray}
It is also numerically suggested to express $\Pi_6$ 
in terms of the 2 Higgs field VEVs as:
\begin{equation}
\Pi_6 = \langle T\rangle^2
\end{equation}
We can now see that it is consistent to split $h_s$ 
in 2 different ways, since we have:
\begin{equation}
h_s = \langle W\rangle\langle T\rangle^2 
= (\underbrace{\langle T\rangle}_{\Pi_2})(\underbrace{\langle W\rangle
\langle T\rangle}_{\Pi_3}) 
= (\underbrace{\langle W\rangle}_{\Pi_4})(\underbrace{\langle T
\rangle^2}_{\Pi_6})
\end{equation}
So we have shown that it may be possible to fit the masses and mixing
angle of the second and third generations 
using the two Higgs fields $W$
and $T$. In particular we make the order of magnitude predictions:
\begin{eqnarray}
h_{\mu} \simeq h_s & \simeq & \langle W\rangle\langle T\rangle^2 
\label{hmuhs} \\
h_c & \simeq & \langle W\rangle^2\langle T\rangle \\
h_{\tau} \simeq h_b & \simeq & \langle W\rangle\langle T\rangle \\
V_{cb} \simeq V_{ts} \simeq V_{23} & \simeq & 
\frac{\langle T\rangle^3}{h_b} 
\simeq \frac{\langle T\rangle^2}{\langle W\rangle}
\label{vcbvts}
\end{eqnarray}

Now we have determined all the $\Pi_i$ and 
so we know the values of $|n_i|$
and $|p_i|$ defined in eq.~(\ref{Pi_i}). 
However, we don't know the signs of
$n_i$ or $p_i$. Essentially, if e.g.\ $n_i<0$ then $|n_i|$ 
hermitean conjugated Higgs fields
$W^{\dagger}$ are used for the corresponding element in the 
mass matrix rather than the Higgs fields $W$ themselves. 
By definition we can choose the sign of the charges
of $W$ and $T$ so that:
\begin{equation}
\vec{b} = \vec{Q}_W + \vec{Q}_T
\label{b=WT}
\end{equation}
Then we can use eq.~(\ref{bcs}), with our reasonable assumption 
that $\vec{Q}_W$ and
$\vec{Q}_T$ are linearly independent, to see that the only 
consistent choice is that
\begin{eqnarray}
\vec{c} & = & -2\vec{Q}_W + \vec{Q}_T  \label{c=-2WT} \\
\vec{s} & = & \vec{Q}_W - 2\vec{Q}_T
\label{s=WTT}
\end{eqnarray}

\section{Predictions from a generalised model}
\label{predictions}

Clearly we can now calculate the charges of 
the Higgs fields $W$ and $T$ in
the AGUT model. However, we shall first 
highlight the predictions that 
only depend on the following general 
features suggested by the AGUT 
model and the data as discussed in section 4:
\begin{enumerate}
\item[(a)] Left-handed up-type and down-type 
quarks have the same charges.
\item[(b)] The ``diagonal'' matrix 
elements in the up-type mass matrix 
involve the opposite charges to those in the corresponding elements 
of the down-type and lepton matrices. The ``diagonal'' elements are 
just meant to be diagonal after some appropriate permutation of the 
imagined left and right handed proto-quarks and leptons.
\item[(c)] The top quark mass has zero quantum numbers suppressing 
it and is ``off-diagonal''.
\end{enumerate}
Now the linear relations, eqs.~(\ref{bcs}) 
and (\ref{bsV}), are in fact 
consequences of these assumptions (a), (b) and (c). Using the 
phenomenological arguments of the previous section, we take as our 
final assumption:
\begin{enumerate}
\item[(d)] The quantum numbers of the 
Higgs fields $W$ and $T$ should be 
related to those for the mass matrix elements by 
eqs.~(\ref{b=WT})-(\ref{s=WTT}).
\end{enumerate}

With the above 4 assumptions extracted from the AGUT model and 
phenomenology, we obtain the Yukawa coupling matrices in the form:
\begin{eqnarray}
H_U & \sim & \left ( \begin{array}{cc} W^{\dagger}T^2 & 
	(W^{\dagger})^2T \\
		1 & W^{\dagger}T^{\dagger} \end{array} \right ) \\
H_D & \sim & \left ( \begin{array}{cc} W(T^{\dagger})^2 & T^3 \\
		W^2(T^{\dagger})^4 & WT \end{array} \right ) \\
H_E & \sim & \left ( \begin{array}{cc} W(T^{\dagger})^2 & ? \\
			? & WT \end{array} \right )
\end{eqnarray}
where the order of magnitudes of the elements 
are obtained by replacing
the Higgs fields by their 
expectation values (in units of $M_F$).
So only the 2 off-diagonal elements in 
the charged lepton matrix are model
dependent. Since we now have all the 
matrix elements for the quarks we can
check that we obtain the required masses 
and mixing angle, i.e.\ that no
elements dominate the ones we expected to 
be relevant for calculating the
eigenvalues.

The largest eigenvalue in a $2 \times 2$ 
matrix is approximately the largest
element and the other eigenvalue is 
approximately the determinant divided by
the largest element. So we can see that 
the up-type matrix has the form we
expected, since no element is larger than 1 and
$\langle W\rangle^2\langle T\rangle \gg 
(\langle W\rangle\langle T\rangle^2)(\langle W\rangle\langle T\rangle)$. 
In the down-type matrix we see that the
element we wanted to be $h_bV_{23}$ has the correct magnitude 
$\langle T\rangle^3$. We also
see that the other off-diagonal element 
in $H_D$ is greatly suppressed and does not
give any significant contribution to the eigenvalues.

As already discussed in section 4, the first generation $u$, $d$ and 
$e$ masses could be dominated by the extra diagonal elements in the 
3 generation Yukawa coupling matrices $H_U$, $H_D$ and $H_E$. 

It should be noted that when we 
consider a specific model such as the AGUT
model, problems can arise. 
Most obviously the off-diagonal elements 
in $H_E$ could
dominate one or both of the required 
eigenvalues. In this case the model
would not be suitable. Secondly, it might turn out 
that the charges of the
Higgs fields $W$ and $T$ were not independent and then 
the elements would be expressible as different combinations of these
fields. In this case the matrix 
element would correspond to the least
suppressed combination and this may 
be different from the elements suggested
above. Finally, a similar situation 
arises when elements involving the
first generation are considered. 
Then it may be necessary (as it is in fact
in the AGUT model) to introduce more Higgs fields, since all the
elements may not be expressible in terms of $W$ and $T$ alone. 
If all the Higgs fields are
not linearly independent then we again have 
the possibility that there are
other less suppressed combinations. These last two situations do not
necessarily spoil the model, but a careful check is required.

\section{Specific choice of Higgs fields within the AGUT model}
\label{higgs}

We are now ready to calculate the charges of 
the Higgs fields $W$ and $T$ in
the AGUT model. This is quite simple using eqs.~(\ref{b=WT}) and
(\ref{c=-2WT}). We obtain:
\begin{eqnarray}
\vec{Q}_W = \frac{1}{3}(\vec{b}-\vec{c}) & = &
		\left (0,-\frac{1}{2},\frac{1}{2},-\frac{4}{3} \right ) 
\label{qw}\\
\vec{Q}_T = \vec{b}-\vec{Q}_W & = &
		\left (0,-\frac{1}{6},\frac{1}{6},-\frac{2}{3} \right )
\label{qt}
\end{eqnarray}
Obviously these charges are linearly independent 
and so the second potential
problem, discussed at the end of the previous section, 
does not arise in this model.
We can now calculate all the elements involving 
only the second and third
generations. The relevant elements of the three 
Yukawa matrices are given by:
\begin{eqnarray}
H_U & \sim & \left ( \begin{array}{cc} W^{\dagger}T^2 & 
		(W^{\dagger})^2T \\
		1 & W^{\dagger}T^{\dagger} \end{array} \right ) \\
H_D & \sim & \left ( \begin{array}{cc} W(T^{\dagger})^2 & T^3 \\
		W^2(T^{\dagger})^4 & WT \end{array} \right ) \\
H_E & \sim & \left ( \begin{array}{cc} W(T^{\dagger})^2 & 
		W^3(T^{\dagger})^5 \\
		(W^{\dagger})^2T^4 & WT \end{array} \right )
\end{eqnarray}
Now it can clearly be seen that the off-diagonal 
elements in the charged
lepton matrix are sufficiently suppressed, so that they do not make
significant contributions to the eigenvalues. 
Thus, as required, the first
potential problem discussed at the end of the previous section 
does not arise in this model. So we
have shown that this model could give 
realistic fermion masses for the
second and third generations. We must 
now complete the model by determining
the matrix elements involving the first generation.

We notice that the charges of $W$ and $T$ 
do not cover the 2 dimensional space
of charges $\frac{y_3}{2}$ and $y_f$, 
since only even $y_f$ charges can be
constructed with integer numbers of these Higgs fields.
Therefore, since both $W$
and $T$ have $\frac{y_1}{2}=0$, we will
need at least 2 more Higgs fields to 
fully cover the 3 dimensional charge space
required to break G down to the SMG. We will now choose 2
more Higgs fields which, together with $W$ and $T$,
will fully cover this space.

As a simple proposal, we may introduce a third Higgs field $\xi$
provided with charges corresponding to the quantum number differences 
between the left-handed quarks in the first and second generations.
Then the  $H_U$ and $H_D$ matrix elements in the row corresponding 
to the left-handed first generation quark become the same as those 
in the second generation row augmented by an extra factor $\xi$.
The $V_{us}$ mixing matrix element is then expected to be dominated 
by the $H_U$ and $H_D$ elements in the first generation row, 
corresponding to those in the second row which dominate the $c$ and 
$s$ masses. In fact $V_{us}$ is readily seen to be given, in our type 
of model, by the ratio of these first to second row matrix elements 
and should thus be equal order of magnitudewise to the $\xi$ 
Higgs field suppression factor:  
\begin{equation}
V_{us} \simeq \langle \xi\rangle \simeq 0.22
\label{Vus}
\end{equation}
Similarly we expect:
\begin{equation}
V_{ub} \simeq \langle \xi\rangle V_{23} \simeq V_{us}V_{cb}
\end{equation}
The quantum numbers for such a $\xi$ field must of course be:
\begin{equation}
\vec{Q}_{\xi} = \vec{Q}_{d_L} - \vec{Q}_{s_L}
        = \left( \frac{1}{6},0,0,0 \right) - 
	\left( 0,\frac{1}{6},0,0 \right)
        = \left( \frac{1}{6},-\frac{1}{6},0,0 \right)
\label{qxi}
\end{equation}

We must now choose one more Higgs field to fully span
the 3 dimensional space of charges. 
Otherwise the first generation would 
remain massless. In order to be consistent with the well-known 
phenomenological relation:
\begin{equation}
V_{us} \simeq \sqrt{\frac{m_d}{m_s}}
\end{equation}
which motivated the Fritzsch \cite{Fritzsch} and many subsequent 
quark mass ans\"{a}tze, we clearly want the first generation 
``diagonal" matrix elements to be suppressed, 
by a factor $\langle \xi\rangle^2$ 
relative to the Yukawa matrix element dominating the $s$ quark mass. 
With our proposal that the first generation row has a factor 
$\xi$ more than the second generation row, we must thus arrange 
for the transition matrix element from $s_L$ to $d_R$ to have 
the same factor $W(T^{\dagger})^2\xi$ as that from $d_L$ to $s_R$. 
This can be achieved---numerically---by proposing a Higgs field $S$,
having a suppression factor that is 
actually equal to unity and that can 
compensate for the difference in quantum numbers between these two 
hoped for equally suppressed matrix elements.
This will lead to 2 different but comparable mechanisms for the down
quark mass. So we take:
\begin{equation}
\langle S\rangle = 1
\end{equation}
and the charges of $S$ are given by:
\begin{eqnarray}
\vec{Q}_{S} & = & [\vec{Q}_{s_L} - \vec{Q}_{d_R}]
                - [\vec{Q}_{d_L} - \vec{Q}_{s_R}] \nonumber \\
 & = & \left[ \left( 0,\frac{1}{6},0,0 \right) -
                \left( -\frac{1}{3},0,0,0 \right) \right] -
        \left[ \left( \frac{1}{6},0,0,0 \right) -
                \left( 0,-\frac{1}{3},0,-1 \right) \right] \nonumber \\
 & = & \left( \frac{1}{6},-\frac{1}{6},0,-1 \right)
\label{qs}
\end{eqnarray}
We note that the SM weak hypercharge vanishes
\begin{equation}
y = y_1 + y_2 + y_3 = 0
\end{equation}
for the Higgs fields $W$, $T$, $\xi$ and $S$. This guarantees 
that the SMG is recovered as the diagonal subgroup of the 
$SMG_i$ groups.

We can now calculate the suppression of all 
elements in the Yukawa matrices.
However, we must first note that, since 
we have used 4 Higgs fields, we cannot
uniquely resolve the charge differences 
between left-handed and right-handed
fermions. There will be some combination of the 4 Higgs field
charges which will result in vanishing charge differences.
We must find the smallest combination of
the 4 Higgs fields which results in a
vanishing set of charges $\vec{Q}=0$. To do this we note that all
fermion ${\rm U}(1)_f$ charge differences 
are quantised as integers. 
However, the 3
Higgs fields $W$, $T$ and $\xi$ can 
only give integer ${\rm U}(1)_f$
charge differences which are even.
Therefore we must have at least two $S$ Higgs fields
involved in the combination. Then we can find
the simplest combination of the other 3 Higgs fields which,
together with the two $S$ fields, give
net vanishing charge differences $\vec{Q}=0$.
This combination is:
\begin{equation}
2\vec{Q}_{S}-2\vec{Q}_{\xi}-9\vec{Q}_{T}+3\vec{Q}_{W} = 0
\end{equation}
Since this involves such large powers
of $T$, there is usually no ambiguity in selecting the combination
of Higgs fields which suppresses the transition 
the least. So finally we can
write out the full Yukawa matrices for each type of fermion:
\begin{eqnarray}
H_U & \sim & \left ( \begin{array}{ccc}
	S^{\dagger}W^{\dagger}T^2(\xi^{\dagger})^2 & W^{\dagger}T^2\xi &
		(W^{\dagger})^2T\xi \\
	S^{\dagger}W^{\dagger}T^2(\xi^{\dagger})^3 & W^{\dagger}T^2 &
		(W^{\dagger})^2T \\
	S^{\dagger}(\xi^{\dagger})^3 & 1 & W^{\dagger}T^{\dagger}
			\end{array} \right ) \label{H_U} \\
H_D & \sim & \left ( \begin{array}{ccc}
	SW(T^{\dagger})^2\xi^2 & W(T^{\dagger})^2\xi & T^3\xi \\
	SW(T^{\dagger})^2\xi & W(T^{\dagger})^2 & T^3 \\
	SW^2(T^{\dagger})^4\xi & W^2(T^{\dagger})^4 & WT
			\end{array} \right ) \label{H_D} \\
H_E & \sim & \left ( \begin{array}{ccc}
	SW(T^{\dagger})^2\xi^2 & W(T^{\dagger})^2(\xi^{\dagger})^3 &
		(S^{\dagger})^2WT^4\xi^{\dagger} \\
	SW(T^{\dagger})^2\xi^5 & W(T^{\dagger})^2 & 
	(S^{\dagger})^2WT^4\xi^2 \\
	S^3W(T^{\dagger})^5\xi^3 & (W^{\dagger})^2T^4 & WT
			\end{array} \right ) \label{H_E}
\end{eqnarray}
Note that one of the elements in $H_E$ has been 
changed after the introduction
of the Higgs fields $\xi$ and $S$, 
but this element is still so suppressed that
it has practically no relevance for any of 
the charged lepton masses.

\section{Results}
\label{results}

Now we are able to choose specific values for the 3 VEVs 
$\langle W\rangle$, $\langle T\rangle$ 
and $\langle \xi\rangle$ and calculate
the resulting masses and mixing angles. 
The overall mass scale for the 
fit is set by eqs.~(\ref{mass-scale}) and (\ref{WS-vev}). 
In order to find the best possible fit we
must use some function which measures how 
good a fit is. Since we are expecting
an order of magnitude fit, this function 
should depend only on the ratios of
the fitted masses to the experimentally 
determined masses. The obvious choice
for such a function is:
\begin{equation}
\chi^2=\sum \left[\ln \left(
\frac{m}{m_{\mbox{\small{exp}}}} \right) \right]^2
\end{equation}
where $m$ are the fitted masses and mixing angles and
$m_{\mbox{\small{exp}}}$ are the
corresponding experimental values. The Yukawa
matrices are calculated at the fundamental scale 
which we take to be the
Planck scale. We use the first order renormalisation 
group equations (RGEs) for
the SM to calculate the matrices at lower scales.
Running masses are calculated in terms of the Yukawa 
couplings at 1 GeV.
The only exception is the top quark, 
where the experimentally measured mass is
the pole mass and this is what we quote. 
We present here the result
of an updated fit, using the following 
values \cite{PDG} of the SM gauge 
coupling constants at the electroweak scale and their values 
extrapolated to the Planck scale via the RGEs:
\begin{eqnarray}
U(1):   \qquad & g_1(M_Z) = 0.462 \quad & g_1(M_{Planck}) = 0.614\\
SU(2):  \qquad & g_2(M_Z) = 0.651 \quad & g_2(M_{Planck}) = 0.504\\
SU(3):  \qquad & g_3(M_Z) = 1.22  \quad & g_1(M_{Planck}) = 0.491
\end{eqnarray}
We cannot simply use the 3 matrices given by
eqs.~(\ref{H_U})--(\ref{H_E}) to calculate 
the masses and mixing angles, since
only the order of magnitude of the elements is defined. 
This could result in
accidental cancellations if we calculated the 
eigenvalues and eigenvectors
using these values. Therefore we calculate 
statistically, by giving each
element a random complex phase and then 
finding the masses and mixing angles.
We repeat this several times and calculate 
the geometrical mean
for each mass and mixing
angle. In fact we also vary the magnitude 
of each element randomly, by
multiplying by a factor chosen to be 
the exponential of a number picked from a
Gaussian distribution with mean value 0 and standard deviation 1.

We then vary the 3 free parameters to 
find the best fit given by the $\chi^2$
function. We get the lowest value of $\chi^2$ for the VEVs:
\begin{eqnarray}
\langle W\rangle & = & 0.179   \label{Wvev} \\
\langle T\rangle & = & 0.071   \label{Tvev} \\
\langle \xi\rangle & = & 0.099 \label{xivev}
\end{eqnarray}
The fitted value of $\langle \xi\rangle$ is approximately 
a factor of two smaller than the
estimate given in eq.~(\ref{Vus}). 
This is mainly because there are
contributions to $V_{us}$ of the same 
order of magnitude from both $H_U$ and
$H_D$. The result of the fit is shown 
in table~\ref{convbestfit}. The
experimental values are those given 
in table~\ref{ExpMasses}. This fit has a
value of:
\begin{equation}
\chi^2=1.87
\label{chisquared}
\end{equation}
This is equivalent to fitting 9 degrees of
freedom (9 masses + 3 mixing angles - 3
Higgs VEVs) to within a factor of 
$\exp(\sqrt{1.87/9}) \simeq 1.58$
of the experimental value. This is 
better than would have been
expected from an order of magnitude 
fit and should be compared with
$\chi^2=3.7$ for the fit with only 7 degrees 
of freedom in \cite{Gerry2}.

\begin{table}
\caption{Best fit to conventional experimental data. 
All masses are running
masses at 1 GeV except the top quark mass which is the pole mass.}
\begin{displaymath}
\begin{array}{ccc}
\hline
 & {\rm Fitted} & {\rm Experimental} \\ \hline
m_u & 3.6 {\rm \; MeV} & 4 {\rm \; MeV} \\
m_d & 7.0 {\rm \; MeV} & 9 {\rm \; MeV} \\
m_e & 0.87 {\rm \; MeV} & 0.5 {\rm \; MeV} \\
m_c & 1.02 {\rm \; GeV} & 1.4 {\rm \; GeV} \\
m_s & 400 {\rm \; MeV} & 200 {\rm \; MeV} \\
m_{\mu} & 88 {\rm \; MeV} & 105 {\rm \; MeV} \\
M_t & 192 {\rm \; GeV} & 180 {\rm \; GeV} \\
m_b & 8.3 {\rm \; GeV} & 6.3 {\rm \; GeV} \\
m_{\tau} & 1.27 {\rm \; GeV} & 1.78 {\rm \; GeV} \\
V_{us} & 0.18 & 0.22 \\
V_{cb} & 0.018 & 0.041 \\
V_{ub} & 0.0039 & 0.0035 \\ \hline
\end{array}
\end{displaymath}
\label{convbestfit}
\end{table}

We can also fit to different experimental values 
of the 3 light quark
masses by using recent results from lattice QCD
\cite{udsMasses}. Light quark masses extracted from lattice 
QCD seem to be consistently lower than the conventional
phenomenological values \cite{PDG} given in table~\ref{ExpMasses}.
We take the following light quark masses as typical lattice values,
extrapolated to 1 GeV using the RGEs:

\begin{eqnarray}
m_u & \simeq & 1.3 \; MeV \\
m_d & \simeq & 4.2 \; MeV \\
m_s & \simeq & 85 \; MeV
\end{eqnarray}
We can now vary the Higgs VEVs to give the best 
fit to this alternative data.
The best fit is shown in table~\ref{newbestfit}. 
The values of the Higgs VEVs are:
\begin{eqnarray}
\langle W\rangle & = & 0.123	\\
\langle T\rangle & = & 0.079	\\
\langle \xi\rangle & = & 0.077
\end{eqnarray}
and this fit has a larger value of:
\begin{equation}
\chi^2 = 3.81
\end{equation}

\begin{table}
\caption{Best fit using alternative light quark masses extracted from
lattice QCD. All masses are running
masses at 1 GeV except the top quark mass which is the pole mass.}
\begin{displaymath}
\begin{array}{ccc}
\hline
 & {\rm Fitted} & {\rm Experimental} \\ \hline
m_u & 1.9 {\rm \; MeV} & 1.3 {\rm \; MeV} \\
m_d & 3.7 {\rm \; MeV} & 4.2 {\rm \; MeV} \\
m_e & 0.45 {\rm \; MeV} & 0.5 {\rm \; MeV} \\
m_c & 0.53 {\rm \; GeV} & 1.4 {\rm \; GeV} \\
m_s & 327 {\rm \; MeV} & 85 {\rm \; MeV} \\
m_{\mu} & 75 {\rm \; MeV} & 105 {\rm \; MeV} \\
M_t & 192 {\rm \; GeV} & 180 {\rm \; GeV} \\
m_b & 6.4 {\rm \; GeV} & 6.3 {\rm \; GeV} \\
m_{\tau} & 0.98 {\rm \; GeV} & 1.78 {\rm \; GeV} \\
V_{us} & 0.15 & 0.22 \\
V_{cb} & 0.033 & 0.041 \\
V_{ub} & 0.0054 & 0.0035 \\ \hline
\end{array}
\end{displaymath}
\label{newbestfit}
\end{table}

Comparing this fit to the one above, 
using conventional light quark masses,
we can see that we have an improvement in
the first generation masses, since the up and 
down quark masses were lowered towards
the electron mass. The fit to $V_{cb}$ has also improved. 
However, the strange quark mass, which we always predicted too
large, is even worse because it's 
experimental mass has also been lowered.
It may seem that
now our assumption, that $h_s \simeq h_{\mu}$ 
at the fundamental scale, is
not correct. Of course we must remember that 
everything should be taken order
of magnitudewise 
and we can always ignore one borderline case such as this.
However, it would be interesting to see 
if we could have produced a model
without the order of magnitude degeneracy between the 
strange quark and and muon masses.

\section{Suggestion for a better model of the alternative masses}
\label{alternative}

At first it would appear to be impossible 
to produce a model without the order of magnitude degeneracy 
$h_s \simeq h_{\mu}$, while retaining the other order
of magnitude degeneracies between the leptons 
and down-type quarks. Indeed,
we would certainly not want to spoil the natural prediction of
$h_b \simeq h_{\tau}$, which is well-known 
to be quite accurate. The relation
between the down and electron masses may 
not appear to be so accurate.
However, when we consider the lower 
value of the down quark mass extracted from lattice QCD, 
along with the hindsight realisation that 
we actually predict $h_d \simeq 2h_e$ due
to the two competing combinations of 
elements giving the lowest eigenvalue,
we actually have quite 
an accurate prediction which we would not want to spoil.
As we noted in section~\ref{boot}, if one eigenvalue 
involves an off-diagonal term, then so must 
at least one other. Thus, it would appear that we cannot spoil
the unwanted relation, $h_s \simeq h_{\mu}$, 
without spoiling another relation.

However, we can in fact do precisely this. 
Examining the Yukawa matrices
given by eqs.~(\ref{H_D}) and (\ref{H_E}), 
we see that there is one
off-diagonal element which has the same 
order of magnitude in the down-type
and charged lepton matrices. It turns out 
that this is not by chance and is
actually a consequence of the element 
being in the same position as the
unsuppressed element in the up-type matrix, 
which leads to the top quark mass. To see
this in general, we use the 
notation of section~\ref{AGUT} for the net charge
differences supplied by the Higgs fields suppressing 
each element in the Yukawa matrices.

If the top mass is to be unsuppressed then 
by definition we have, for some
$i$ and $j$:
\begin{equation}
\Delta\vec{Q}_{Uij} = \vec{Q}_{U_{iL}} - \vec{Q}_{U_{jR}} 
= \vec{Q}_{D_{iL}} - \vec{Q}_{U_{jR}} = \vec{0}
\label{DQt=0}
\end{equation}
We now wish to show that, for the same fixed 
$i$ and $j$, this implies:
\begin{equation}
\Delta\vec{Q}_{Dij} = -\Delta\vec{Q}_{Eij}
\label{DQD=-DQE}
\end{equation}
where we know the sign is minus (either sign would 
lead to the order of
magnitude equality) because of the 
specific example of eqs.~(\ref{H_D}) and
(\ref{H_E}). So now we can see that, using eq.~(\ref{DQt=0}):
\begin{eqnarray}
\Delta\vec{Q}_{Dij} & = & 2\vec{Q}_{D_{iL}} - \vec{Q}_{D_{jR}}
				- \vec{Q}_{U_{jR}} \\
-\Delta\vec{Q}_{Eij} & = & -\vec{Q}_{D_{iL}} - \vec{Q}_{E_{iL}}
				+ \vec{Q}_{U_{jR}} + \vec{Q}_{E_{jR}}
\end{eqnarray}

We can see from table~\ref{Q_f} that all left-handed fields 
carry zero ${\rm U}(1)_f$ charge and so:
\begin{equation}
2\vec{Q}_{D_{iL}} = -\vec{Q}_{D_{iL}} - \vec{Q}_{E_{iL}}
\label{yanom}
\end{equation}
since the only non-zero charges are the 
${\rm U}(1)_i$ charges, which are
equal to the weak hypercharges in the SM. 
An alternative, perhaps more
fundamental, reason for this equality is 
that it corresponds to the absence
of anomalies. In particular the fields 
$D_{iL}$ (which is the same as $U_{iL}$)
and $E_{iL}$ are the only fields coupling 
to ${\rm SU}(2)_i$. So eq.~(\ref{yanom}) is simply
the condition for cancellation of 
anomalies associated with the triangle
Feynman diagrams with two external 
${\rm SU}(2)_i$ gauge bosons and one
U(1) gauge boson (with 4 independent choices of 
the U(1) gauge group).

Similarly the relation:
\begin{equation}
-\vec{Q}_{D_{jR}} - \vec{Q}_{U_{jR}} = \vec{Q}_{U_{jR}} 
+ \vec{Q}_{E_{jR}}
\end{equation}
can be seen to be true. Again this is 
due to anomaly cancellation, though not
in such a simple manner as the cancellation 
for the left-handed fields.
So we now see that eq.~(\ref{DQD=-DQE}) is true. 
Therefore it may be possible
to produce a model, where the bottom quark 
and tau lepton masses come from
the element in the same off-diagonal position 
as the unsuppressed element in
the up-type matrix. This would retain the 
good relation, $h_b \simeq h_{\tau}$,
as well as the order of magnitude 
degeneracy of the 1st generation masses, but
would not enforce the less desirable prediction 
of $h_s \simeq h_{\mu}$. This
interesting scenario is currently 
being investigated and will not be further
commented on in this paper.

\section{CP violation}
\label{CP}

Another prediction, which can be made from a model of 
the mass matrices, is the
amount of CP violation due to the CKM matrix. 
This depends on the complex
phases in the matrix usually parameterised 
by a phase $\delta$ \cite{PDG}.
However, there are different ways of 
parameterising the unitary CKM matrix,
and so it is better to define a 
parameterisation-independent quantity which
is a measure of the amount of CP violation. 
Such a definition is possible and
corresponds to the areas of the ``unitarity triangles'' 
\cite{Jarlskog}. The
3 sides of the triangles are defined in the complex plane as
$s_i = V_{ij}V_{ik}^*$,
where $j \ne k$ are fixed and $i$ labels the 3 sides. The
condition that the CKM matrix $V$ is unitary determines that:
\begin{equation}
s_1 + s_2 + s_3 = 0
\end{equation}
and so these 3 lines in
the complex plane form a triangle. Also, 
the areas of all the different
triangles are the same. Therefore we can 
define the amount of CP violation,
$J$, in terms of the area, $A$, of any of these triangles, e.g.:
\begin{equation}
J = 2A = |x_1y_2-x_2y_1|
\end{equation}
where $x_i$ ($y_i$) are the real (imaginary) components of $s_i$.
In our model with the conventional values of the 
light quark masses, we find
that the model predicts:
\begin{equation}
J \simeq 5.8 \times 10^{-6}
\label{J-fit1}
\end{equation}
In the case where we use the lower lattice values for the light 
quark masses, we have the prediction:
\begin{equation}
J \simeq 1.2 \times 10^{-5}
\label{J-fit2}
\end{equation}

These predicted values are both lower than the experimental 
determination \cite{ExpCP}:
\begin{equation}
J \simeq 2.0 \times 10^{-5} - 3.5 \times 10^{-5}
\label{J-exp}
\end{equation}
The model predictions agree in order of magnitude though, for the
case where we take the conventional values 
of the light quark masses, the
prediction is clearly worse than any of the mass and
mixing angle predictions.
However, we should really even consider this worst case as 
sufficiently good agreement with experiment, since our prediction 
of the CP violation parameter J is expected to be less accurate than 
that for a typical mass or mixing angle. 
This is because our prediction for $J \sim \frac{T^4\xi^2}{W^2}$ 
involves 8 Higgs field VEVs, whereas a typical mass prediction
involves rather of the order of 3 VEVs such as $m_s \sim WT^2$.
As noted after eq.~(\ref{chisquared}), with $\chi^2 = 1.87$, we
expect to fit a typical mass within a (one ``standard deviation'') 
factor of $\exp(\sqrt{1.87/9}) \simeq 1.58 = \exp(\pm 0.46)$.
If now the logarithm of the uncertainty factor for J is
taken to be $\frac{8}{3}$ times as big, then J is 
expected to be uncertain by a factor of $\exp(\pm 1.2) = 3.4^{\pm 1}$.
So even the worst case, eq.~(\ref{J-fit1}), only deviates from
the experimental value, eq.~(\ref{J-exp}), by a factor of
$\log\frac{2.7 \times 10^{-5}}{5.8 \times 10^{-6}}/1.2 = 1.3$ 
``standard deviations''. Thus we conclude that our CP violation 
predictions agree with experiment, within the accuracy we 
can expect.

We will now show that the amount of CP violation in our model can be
well estimated
in terms of the fitted mixing angles. We will first consider the
simplified case, where the up-type Yukawa 
matrix is approximately diagonal, and
so all the quark mixing is due to off-diagonal 
elements in the down-type
Yukawa matrix. Then the down-type matrix should be of the form:
\begin{equation}
H_d \sim \left ( \begin{array}{ccc} h_d & h_sV_{12} & z \\
				      x & h_s & h_bV_{23} \\
				      x & x & h_b \end{array} \right )
\end{equation}
where $x$ denotes element which are considered to be ``sufficiently
suppressed''. This means that the lower 
off-diagonal components should
not be so large as to be relevant when 
calculating masses and mixing angles.
Since the matrix diagonalised is $H_dH_d^{\dagger}$, 
it can easily be seen
that these elements will not be relevant 
unless they are considerably
larger than the corresponding upper off-diagonal elements. This is
because they get multiplied by smaller diagonal elements, since
$h_d \ll h_s \ll h_b$. As we will show:
\begin{eqnarray}
V_{us} \simeq V_{cd} \simeq V_{12} & & \\
V_{cb} \simeq V_{ts} \simeq V_{23}
\end{eqnarray}
However, the element $z$ determines $V_{ub}$ but only contributes 
to $V_{td}$. If $z$ is bigger than or of the same order of magnitude 
as the product 
$h_bV_{us}V_{cb}$,
we would expect that we should have:
\begin{equation}
z \simeq h_bV_{ub} \simeq h_bV_{td}
\end{equation}
but if $z$ were smaller than this value, 
we would still predict the correct
order of magnitude of $V_{td}$ through 
the `indirect' mixing between all
3 generations as opposed to the `direct' 
mixing due to $z$, as we shall now explain.

Consider how to approximately diagonalise 
the matrix $H_dH_d^{\dagger}$.
If $z$ is so small that there is essentially 
no `direct' contribution to
$V_{td}$, then we can use a 2 stage process 
using unitary matrices of the
form:
\begin{equation}
U = \left ( \begin{array}{cc} U_{2 \times 2} & 
\begin{array}{c} 0 \\ 0
\end{array} \\ \begin{array}{cc} 0 & 0 \end{array} & 
1 \end{array} \right )
\; {\rm and} \;
V = \left ( \begin{array}{cc} 1 & \begin{array}{cc} 0 & 
0 \end{array} \\
\begin{array}{c} 0 \\ 0 \end{array} & 
V_{2 \times 2} \end{array} \right ).
\end{equation}
where, in the approximation of small mixing, we have the order of
magnitude unitary matrices (suppressing all phases for convenience):
\begin{equation}
U_{2 \times 2} \simeq \left ( \begin{array}{cc} 1 & V_{12} \\
		V_{12} & 1 \\ \end{array} \right )
\; {\rm and} \;
V_{2 \times 2} \simeq \left ( \begin{array}{cc} 1 & V_{23} \\
                V_{23} & 1 \\ \end{array} \right )
\end{equation}
The order these matrices are applied is important. It can be seen that 
when $z$ is small:
\begin{equation}
U^{\dagger}V^{\dagger}Y_dY_d^{\dagger}VU \simeq {\rm 
diag}(y_d^2,y_s^2,y_b^2).
\end{equation}
If we had applied the matrices in the opposite order we would not have 
approximately diagonalised $Y_dY_d^{\dagger}$. Therefore the CKM matrix 
is given by the product $VU$. With the approximations we have made 
above, this is:
\begin{equation}
V_{\rm CKM} \simeq VU \simeq
		\left ( \begin{array}{ccc} 1 & V_{12} & 0 \\
			V_{12} & 1 & V_{23} \\
			V_{12}V_{23} & V_{23} & 1 \end{array} \right )
\end{equation}
So we see that we have the symmetrical relations:
\begin{eqnarray}
V_{cd} & \simeq & V_{us} \simeq V_{12} \\
V_{ts} & \simeq & V_{cb} \simeq V_{23}
\end{eqnarray}
but the mixing between 1st and 3rd generations is given by:
\begin{eqnarray}
V_{ub} & \simeq & 0 \\
V_{td} & \simeq & V_{12}V_{23} \equiv V_{13}^{\rm indirect}
\end{eqnarray}
This is what we refer to as `indirect' mixing 
between the 1st and 3rd
generations.

Now we shall consider the case when there 
is also `direct' mixing between the 1st
and 3rd generations. This is the case where:
\begin{equation}
z \simeq h_bV_{13}^{\rm direct}
\end{equation}
In this case we can approximately 
diagonalise $H_dH_d^{\dagger}$ by using
the unitary matrix:
\begin{equation}
W = \left ( \begin{array}{ccc} 1 & 0 & V_{13}^{\rm direct} \\
			0 & 1 & 0 \\
			V_{13}^{\rm direct} & 0 & 1 \end{array} \right )
\end{equation}
and then applying the matrices 
$V$ and $U$ as above. In the approximation of 
small mixing angles, this
will only affect the mixing between 
1st and 3rd generations. We will get:
\begin{eqnarray}
V_{ub} & \simeq & V_{13}^{\rm direct} \\
V_{td} & \simeq & V_{13}^{\rm direct} + V_{13}^{\rm indirect}
\end{eqnarray}
So we have found the relation:
\begin{equation}
V_{td} \simeq V_{ub} + V_{us}V_{cb}
\end{equation}
In fact these 3 terms correspond to the 3 sides of one 
of the ``unitarity triangles''. 
The `direct' and `indirect' terms arise from different matrix 
elements of $H_d$, which have been assigned independent random
phases that are averaged over.
This means that, for given values 
for the magnitudes of $V_{ub}$ and
$V_{us}V_{cb}$, the angle $\theta$ between 
these 2 sides should be random.
We can then estimate the amount of 
CP violation (which is two times the area
of this triangle) to be:
\begin{equation}
J \simeq \frac{1}{2}V_{us}V_{cb}V_{ub}
\label{J}
\end{equation}
where the $\frac{1}{2}$ is the geometric average of $\sin\theta$.

Actually there is a contribution to the 
CKM matrix from the up-type
Yukawa matrix in our model and thus 
the above considerations are not quite 
true, in the sense that we do not really have then that the 
phases of two of the sides in the 
unitarity triangle are independent.
In fact the combination of the up-type 
12-transition matrix element and the
23-element in the down-type matrix leads 
to a contribution to the 
CKM matrix, which in turn gives rise to contributions to 
the $V_{ub}$ and the $V_{12}V_{23}$ 
sides of the unitarity triangle.
By a similar calculation as above, 
and with obvious notation, we have:
\begin{eqnarray}
V_{td} & \simeq & V_{13}^{\rm direct} + V_{12}^{\rm down}V_{23} \\
V_{ub} & \simeq & V_{13}^{\rm direct} + V_{12}^{\rm up}V_{23} \\
V_{cb} & \simeq & V_{23} \\
V_{us} & \simeq & V_{12}^{\rm down} + V_{12}^{\rm up}
\end{eqnarray}
The phases are such that $V_{td}$, $V_{ub}$ 
and $V_{us}V_{cb}$ are the 
three sides of a unitarity triangle. 
It is now clear that there are three 
independent quantities relevant 
to the triangle: $V_{13}^{\rm direct}$, 
$V_{12}^{\rm down}V_{23}$ 
and $V_{12}^{\rm up}V_{23}$. Furthermore, in 
our model these three quantities 
are of the same order of magnitude. 
Therefore there is a permutation symmetry 
among the three sides of the 
unitarity triangle.
With such a symmetry it is impossible for
the angles to be flatly distributed, since each angle must have 
an expectation value of $\pi/3$ in order that 
their sum be $\pi$. However,
we expect the distributions to be given by rather smooth functions 
which can naturally be expanded 
on $\cos \theta$, $\cos 2\theta$ etc.
(where the angle is called $\theta$). A contribution of the form
of $\cos( \mbox{``odd''} \theta)$ will 
not influence the average of a function
of $\sin \theta$, such as e.g. $\log \sin \theta$. This 
is because sine and, thus, functions 
of sine take the same value for an angle 
and its supplementary angle, while
$\cos(\mbox{``odd''} \theta)$ changes 
sign between angle and supplementary angle.
So only from even cosines such as $\cos 2\theta$, 
which presumably already comes with a rather
small coefficient, will we expect any modification of the 
average of the sine of the angle. To the accuracy to which this 
$\cos 2\theta$ and higher even cosines 
can be ignored we could, for the
purpose of evaluating the geometric average for $\sin \theta$, 
equally well assume the totally flat 
distribution for $\theta$ used above. 
Thus we do not expect much deviation from
the above estimate, eq.~(\ref{J}), of the CP-violation strength 
by the inclusion of the up-type mass matrix contribution. 
So we can still expect a rather good agreement of 
this estimate with the computer calculation.

We can now calculate our theoretical prediction 
of CP violation, given the
fitted values of the mixing angles for the 2 different fits. 
For the fit
to the conventional light quark masses we expect:
\begin{equation}
J \simeq \frac{1}{2} \times 0.18 \times 0.018 \times 0.0039
			\simeq 6.3 \times 10^{-6}
\end{equation}
and for the fit to the alternative light quark masses:
\begin{equation}
J \simeq \frac{1}{2} \times 0.15 \times 0.033 \times 0.0054
                        \simeq 1.3 \times 10^{-5}
\end{equation}

We can see that the above predictions only deviate by about ten
percent from the calculated values 
eqs.~(\ref{J-fit1}) and (\ref{J-fit2}). 
This deviation is due to the fact
that the up-type Yukawa matrix contributes 
to the mixing between 1st and
2nd generations. However, the theoretical 
prediction still agrees very well
and is a useful estimate in models 
where most of the mixing is due to the
down-type Yukawa matrix and the 
matrix elements have uncorrelated phases.

\section{Neutrino masses and mixing angles}
\label{neutrinos}

We have constructed a successful model of the observed fermion
masses and mixing angles. It is interesting 
to investigate what predictions
this model would give for neutrino masses 
and mixing angles. We expect the
neutrinos to get a mass in this model by the usual see-saw
mechanism \cite{fn2,seesaw}. 
This occurs when we treat the SM as a low energy effective theory. 
Then we can include the non-renormalisable interaction:
\begin{equation}
{\cal L}_{\nu} =
	\frac{1}{M_F}\overline{L_L}\tilde\Phi_{WS}
		H_{\nu}C(\overline{L_L}\tilde\Phi_{WS})^T
			+ {\rm h.c.}
\label{leff}
\end{equation}
with notation as in eq.~(\ref{L_Higgs}), where C is the 
antisymmetric charge conjugation matrix and $H_{\nu}$ 
is the effective light neutrino Majorana coupling matrix. 
After electroweak symmetry breaking
we obtain mass terms for the neutrinos:
\begin{equation}
{\cal L}_{\nu \; {\rm mass}} = \overline{\nu_L}M_{\nu}
C\overline{\nu_L}^T
	+ {\rm h.c.}
\end{equation}
where the Majorana mass matrix:
\begin{equation}
M_{\nu} = H_{\nu}\frac{\langle \phi_{WS}\rangle^2}{2M_F}
\end{equation}
is necessarily symmetric.
Such terms arise in our model in the same way as the 
Yukawa terms for the other
fermions. The only difference is that the Feynman 
diagrams corresponding to
fig.~\ref{MbFull} involve 2 WS Higgs fields 
and the transitions are between
$\nu_{iL}$ and $\nu^c_{jR}$.

The Majorana mass matrix $M_{\nu}$ is, of course, 
a symmetric matrix  
but otherwise, in models of our type with 
approximately conserved chiral 
$U(1)$ charges, the matrix elements 
are generally of different orders 
of magnitude. As pointed out in \cite{fn2}, there are two 
qualitatively different types of eigenstate that can arise 
when diagonalising a symmetric 
mass matrix with such a hierarchical 
structure. In the first case, 
a neutrino can dominantly combine with 
its own antineutrino to form 
a Majorana particle with small mixing 
angles. The second case occurs 
when a neutrino combines dominantly
with an antineutrino, which is 
not the CP conjugate state, to form 
a 2-component massive neutrino. 
For example the bare $\tau$-neutrino 
might combine with the bare $\mu$-antineutrino. Such states 
automatically occur in pairs, with order of magnitudewise 
degenerate masses and maximal mixing. In the example given, 
the other member of the pair would be formed by combining 
the bare $\mu$-neutrino with the bare $\tau$-antineutrino state.
In particular, this happens when the 
matrix element with the largest 
order of magnitude is off-diagonal.

We define the lepton mixing matrix analogously to the CKM
matrix. So we find unitary matrices $U_E$ and $U_{\nu}$ such that:
\begin{eqnarray}
U_E^{\dagger}M_EM_E^{\dagger}U_E & = & \mbox{diag}\{m_e^2, m_{\mu}^2,
m_{\tau}^2\}\\
U_{\nu}^{\dagger}M_{\nu}M_{\nu}^{\dagger}U_{\nu} & = & 
\mbox{diag}\{m_{\nu_e}^2, m_{{\nu}_{\mu}}^2, m_{{\nu}_{\tau}}^2\}
\end{eqnarray}

Then the lepton mixing matrix is defined by:
\begin{equation}
U = U_{\nu}^{\dagger}U_E
\end{equation}

By this mechanism we expect small neutrino masses, 
which may just be large
enough to be observable (if 
the electron neutrino is one member of an almost degenerate 
mass neutrino pair and hence has a large mixing angle) 
as solar neutrino vacuum oscillations. 
Our choice of the Planck scale as the fundamental scale,
$M_F = M_{Planck}$, would require the element 
responsible for the electron
neutrino mass in the Majorana coupling matrix $H_{\nu}$ 
to be essentially unsuppressed (i.e.~of order unity).
In this case 
$m_{\nu_{e}} \sim \frac{\langle \phi_{WS}\rangle^2}{2 M_{Planck}} 
\sim 3 \times 10^{-6}$ eV,
as appropriate for ``just-so'' (energy-dependent) 
vacuum oscillations \cite{just-so}
There is also some experimental evidence for
an oscillation between muon and tau neutrinos. 
This would explain the reduction
in the number of muon neutrinos compared to 
electron neutrinos, observed on
the ground after cosmic ray interactions in the atmosphere.

There are 2 alternative mechanisms for solar neutrino oscillations. 
There is the
obvious possibility of vacuum oscillations, 
i.e. oscillations between the
sun and the earth \cite{vacuum}. 
In order to obtain an oscillation probability which depends on the 
energy of the solar neutrino, as appears necessary when comparing 
the data to the conventional solar 
model calculations \cite{bachall,wark}, 
the neutrinos reaching the earth must have undergone approximately 
one oscillation. The corresponding ``just-so'' 
neutrino mass squared 
difference is $\Delta m^2 \simeq 10^{-11}-10^{-10} \ {\rm eV}^2$. 
The well-known alternative to the large mixing vacuum
oscillation solution is the MSW effect \cite{MSW}. 
This is essentially an enhancement of small mixing oscillations, 
due to electron neutrinos
interacting with electrons within the sun. 
Current solar and atmospheric
neutrino experiments \cite{wark} constrain the 
difference of neutrino masses squared and
mixing angles as shown in table~\ref{NeutrinoData}.

\begin{table}
\caption{Constraints on neutrino mass squared differences 
$\Delta m^2$ and mixing angles $\sin^2 2\theta$ 
from fits \protect\cite{wark} to the solar 
and atmospheric neutrino data.}
\begin{displaymath}
\begin{array}{cccc} \hline
{\rm Experiment} & {\rm Mixing} & \Delta m^2 ({\rm eV}^2)
				& \sin^22\theta \\ \hline
{\rm Solar \; (Vacuum)} & \nu_e - \nu_{\mu,\tau} & 
				10^{-11} - 10^{-10}
				& >0.7 \\
{\rm Solar \; (MSW)} & \nu_e - \nu_{\mu,\tau} & 
				4 \times 10^{-6} - 10^{-5}
				& 3 \times 10^{-3} - 10^{-2} \\
{\rm Atmospheric} & \nu_\mu - \nu_\tau & 
				5 \times 10^{-3} - 3 \times 10^{-2}
				& >0.65 \\ \hline
\end{array}
\end{displaymath}
\label{NeutrinoData}
\end{table}

However, we note that results from recent solar model 
calculations \cite{solarmodel}  
give a predicted $^8B$ neutrino flux which varies by more 
than a factor of two. If the $^8B$ flux is taken to be an 
adjustable parameter within this range, it seems possible to get 
an acceptable energy independent fit to all the solar neutrino 
data \cite{E-independent}. 
By an energy independent solution we mean that
$\Delta m^2$ is sufficiently large
that the ${\nu}_{e} \rightarrow {\nu}_{\mu}$ 
oscillation occurs many times
between the sun and the earth, and what is 
observed is the average probability of
oscillation, which does not depend on the energy of the neutrino.
Such an energy independent large mixing vacuum 
oscillation solution to the solar neutrino 
problem corresponds to a mass squared 
difference $10^{-2} \ {\rm eV}^2 > 
\Delta m^2 > 10^{-10} \ {\rm eV}^2$. The upper limit is 
provided by reactor oscillation experiments \cite{wark}.

We can calculate the neutrino mass matrix in our model, 
if we assume that there
are no new Higgs fields which were not involved 
in the other fermion mass
matrices. In terms of the Higgs fields already introduced 
we find:
\begin{equation}
H_{\nu} \sim \left ( \begin{array}{ccc}
	(S^{\dagger})^2(W^{\dagger})^2T^4(\xi^{\dagger})^4 &
		(S^{\dagger})^2(W^{\dagger})^2T^4\xi^{\dagger} &
		(W^{\dagger})^2T(\xi^{\dagger})^3 \\
	(S^{\dagger})^2(W^{\dagger})^2T^4\xi^{\dagger} &
		W(T^{\dagger})^5 & (W^{\dagger})^2T \\
	(W^{\dagger})^2T(\xi^{\dagger})^3 & (W^{\dagger})^2T &
		S^2(W^{\dagger})^2(T^{\dagger})^2(\xi^{\dagger})^2
			\end{array} \right )
\label{hnu}
\end{equation}
where, as usual, we assume that all fundamental 
Yukawa couplings are of order 1.
Clearly all the elements of $H_{\nu}$ are suppressed. 
The largest element is off-diagonal and of order
$\langle W\rangle^2\langle T\rangle$. As emphasised above, 
the matrix is symmetric 
and so we obtain a pair of almost degenerate eigenvalues. We 
find, using the fitted values eqs.~(\ref{Wvev}-\ref{xivev}),
the following eigenvalues for $H_{\nu}$:
\begin{eqnarray}
h_{\nu_\mu} \simeq h_{\nu_\tau} & \simeq & 
\langle W\rangle^2\langle T\rangle  \simeq   
2.3 \times 10^{-3} \\
h_{\nu_e} & \simeq & \langle W\rangle^2\langle T\rangle^4\langle 
\xi\rangle^4  \simeq 7.8 \times 10^{-11}
\end{eqnarray}

Since a pair of off-diagonal elements dominate the matrix,
there is almost maximal mixing ($\sin^2 2\theta = 1$) 
between $\nu_\mu$ and $\nu_\tau$ with
very small mixing ($\sin^2 2\theta \simeq \langle 
\xi\rangle^6 \simeq 10^{-6}$) of $\nu_e$. This is not suitable 
for observable solar
neutrino oscillations. However, we do have the correct 
mixing structure for
atmospheric neutrino oscillations. The problem here is that 
this would require
the difference in masses squared of $\nu_\mu$ and $\nu_\tau$ to be:
\begin{equation}
\Delta m^2_{\mu\tau} = |m^2_{\nu_{\tau}} - m^2_{\nu_{\mu}}| = 
\Delta m^2_{{\rm exp}} \simeq 10^{-2} \ {\rm eV}^2
\end{equation}
In our model the main contribution to the mass difference, 
between the quasi-degenerate mass eigenstates $\nu_{\tau}$ 
and $\nu_{\mu}$, comes from the largest diagonal element in the 
$2 \times 2$ submatrix of $H_{\nu}$ containing the dominant 
off-diagonal elements. From eq.~(\ref{hnu}), we see that this 
element is $(H_{\nu})_{33}$ and hence:
\begin{equation}
\frac{\Delta m_{\mu\tau}}{m_{\nu_{\tau}}} \simeq 
\frac{(H_{\nu})_{33}}{h_{\nu_{\tau}}} \simeq 
\langle T\rangle\langle \xi\rangle^2  
\simeq 7 \times 10^{-4}
\end{equation}
Unfortunately our scale is totally wrong, since with the natural
choice of the Planck scale as the fundamental scale we predict:
\begin{equation}
m_{\nu_{\mu}} \simeq m_{\nu_{\tau}} \simeq h_{\nu_{\tau}}
\frac{\langle \phi_{WS}\rangle^2}{2M_F} \simeq 
7 \times 10^{-9} \ \mbox{eV}, \ \ m_{\nu_e} 
\simeq 2 \times 10^{-16} \ {\rm eV} 
\end{equation}
and
\begin{equation}
\Delta m^2_{\mu\tau} = 2 \frac{\Delta m_{\mu\tau}}{m_{\nu_{\tau}}}
m^2_{\nu_{\tau}} \simeq 7 \times 10^{-20} \ {\rm eV}^2
\end{equation}

Hence if this model is to generate observable mixing it will be
necessary to introduce a new mass scale 
($M_F \sim 10^{11}\ {\rm GeV}$),
which although both arbitrary and aesthetically unappealing
is usually necessary in models of neutrino mass. Altering
our interpretation of $M_F$ as the Planck scale would spoil the AGUT 
predictions of the gauge coupling constants in \cite{Bennett}, 
and we are reluctant to do so. However, 
it should be noted that an $SMG\otimes U(1)^3$ model with the same
U(1) charges as our AGUT model would be anomaly free. 
Such a model would have the same
fermion mass matrix structure but would not  
suffer from the same objection to a lower $M_F$, as it would 
have no prediction for the values of the gauge coupling constants.
This $SMG\otimes U(1)^3$ model does not appeal so strongly to us, 
precisely because it does not predict values for the gauge 
coupling constants and the choice of the U(1) charges seems 
rather arbitrary \cite{SMG3U1}. 

We may introduce an {\em ad hoc} new mass scale into the AGUT model
and keep $M_F = M_{Planck}$, 
by adding another Higgs particle ($\Delta$ say) 
with an arbitrarily chosen VEV, which is a triplet under SU(2) in
the SM. Then including the interaction:
\begin{equation}
{\cal L}_{\nu} = {\overline{L}}_L \mbox{\boldmath$\tau$}.
\Delta H_{\nu}C
{\overline{L}_L^T}
	+ {\rm h.c.}
\end{equation}
where \mbox{\boldmath${\tau}$} are the Pauli spin matrices,
leads to the neutrino mass matrix:
\begin{equation}
M_{\nu} = H_{\nu}\langle {\Delta}^0\rangle
\end{equation}
Here ${\Delta}^0$ is the neutral component of the Higgs triplet.
The model can now give suitable masses and mixing for the
atmospheric neutrino problem, by the assignment of 
the following quantum numbers, corresponding to the 
exchange of 2 WS Higgs fields, to $\Delta$: 
\begin{equation}
{\vec{Q}}_{\Delta} = -2\vec{Q}_{WS} =
\left(0, -\frac{4}{3}, \frac{1}{3}, -2 \right) 
\end{equation}
It is taken to be a triplet under SU(2) in the $SMG_3$ 
group, but otherwise to belong to singlet or fundamental 
non-Abelian representations of the 
three $SMG_i$ groups. 
With this $\Delta$ we have
the same prediction, eq.~(\ref{hnu}), for $H_{\nu}$, 
but with the choice of a new mass scale 
$\langle {\Delta}^0\rangle \simeq 1300 \ {\rm eV}$
we now have:
\begin{eqnarray}
\Delta m_{\mu\tau}^2 \simeq 10^{-2} \ {\rm eV}^2 \\
m_{{\nu}_{\mu}} \simeq m_{{\nu}_{\tau}} \simeq 3 \ \mbox{eV}, 
\ \ m_{\nu_e} \simeq 10^{-7} \ \mbox{eV} 
\end{eqnarray}
So, as well as explaining the atmospheric neutrino problem, 
the neutrinos $\nu_{\mu}$
and $\nu_{\tau}$ would also be hot dark matter candidates. 
There is also some evidence for 
$\overline{\nu}_{\mu} \rightarrow \overline{\nu}_e$ 
neutrino oscillations from the LSND collaboration \cite{lsnd};
however, as we shall see, we would not 
predict any observable mixing there.
The lepton mixing matrix U for our model is given by:
\begin{eqnarray}
U  \sim \left ( \begin{array}{ccc}
	1 &  \langle \xi\rangle^3 & 
\langle T\rangle^3\langle \xi\rangle \\
	-\frac{\langle \xi\rangle^3}{\sqrt{2}} & \frac{1}{\sqrt{2}}
& \frac{1}{\sqrt{2}} \\
	\frac{\langle \xi\rangle^3}{\sqrt{2}} & -\frac{1}{\sqrt{2}}
& \frac{1}{\sqrt{2}}
			\end{array} \right ) 
\sim \left ( \begin{array}{ccc}
	1 &  10^{-3} & 4 \times 10^{-5} \\
	-7 \times 10^{-4} & 0.71 & -0.71 \\
	7 \times 10^{-4} & 0.71 & 0.71
			\end{array} \right ) \label{U} 
\end{eqnarray}
where we have used the fitted VEVs of eqs.~(\ref{Wvev}-\ref{xivev}) 
and ignored CP violating phases.
Here $U_{l\alpha}$ denotes the mixing between 
flavour eigenstate $l$ and
mass eigenstate $\alpha$.
This leads to the bound:
\begin{eqnarray}
P_{\overline{\nu}_{\mu}\overline{\nu}_e} & \le & \sum_{\alpha,\beta}
\left|U_{\overline{\nu}_e\alpha}U_{\overline{\nu}_e\beta}
U_{\overline{\nu}_{\mu}\alpha}U_{\overline{\nu}_{\mu}\beta}
\right| \nonumber\\ & \simeq & 2 \times 10^{-6}
\end{eqnarray}
on the probability for 
$\overline{\nu}_{\mu} \rightarrow \overline{\nu}_e$
oscillations.
This probability is too small to be observed by current experiments
and, in particular, is incompatible with the LSND 
observation for which:
\begin{equation}
P_{\overline{\nu}_{\mu}\overline{\nu}_e} \simeq 10^{-3}, \qquad
\Delta m^2_{e\mu} \simeq 1 \ {\rm eV}^2
\end{equation}

The above choice of quantum numbers for $\Delta$ is not the only
possibility and, by choosing appropriate charges, it is in fact
possible to explain the solar neutrino problem. 
For example the choice: 
\begin{equation}
\vec{Q}_{\Delta} = \left(-\frac{1}{2}, -\frac{1}{2}, 
0, \frac{1}{3}\right)
\end{equation}
for the Abelian charges, and singlet or fundamental non-Abelian 
representations under the $SMG_i$ groups,
leads to the mass matrix:
\begin{eqnarray}
M_{\nu}  \sim  \langle {\Delta}^0\rangle\left ( \begin{array}{ccc}
        S(T^{\dagger})^3W({\xi}^{\dagger})^4 & 
		S(T^{\dagger})^3W{\xi}^{\dagger} & 
			ST^3(W^{\dagger})^2{\xi}^{\dagger} \\
        S(T^{\dagger})^3W{\xi}^{\dagger} & S(T^{\dagger})^3W{\xi}^2 & 
		S^5T^3(W^{\dagger})^2{\xi}^2 \\
	ST^3(W^{\dagger})^2\xi^{\dagger} & 
	S^5T^3(W^{\dagger})^2{\xi}^2 & 
		S^3(W^{\dagger})^2
                        \end{array} \right ) 
\label{mnu}
\end{eqnarray}
In this case the largest element $(M_{\nu})_{33}$ 
is diagonal and corresponds to the mass of
$\nu_{\tau}$, while the masses of the lighter $\nu_{\mu}$ and 
$\nu_e$ states are quasi-degenerate corresponding to the off-diagonal 
elements $(M_{\nu})_{12} =  (M_{\nu})_{21}$. For
$\langle {\Delta}^0\rangle \simeq  3 \ \mbox{eV}$
we find that 
\begin{eqnarray}
\Delta m_{e\mu}^2 \simeq 7 \times 10^{-11} \ {\rm eV}^2\\
m_{{\nu}_e} \simeq m_{{\nu}_{\mu}} \simeq 2 \times 10^{-5} \ \mbox{eV}, 
\ \ m_{\nu_{\tau}} \simeq 10^{-1} \ \mbox{eV} 
\end{eqnarray}
which, since we have almost maximal mixing between ${\nu}_e$ and
${\nu}_{\mu}$, is suitable for the ``just-so'' 
vacuum oscillation solution to the solar neutrino problem given 
in table~\ref{NeutrinoData}.

As we noted earlier, the allowed range of values for ${\sin}^22\theta$
and $\Delta m^2$ from table~\ref{NeutrinoData} is for energy dependent
solutions of the solar neutrino problem.
However it is also possible to get energy independent solutions 
with maximal mixing and 
$\Delta m^2_{e\mu} > 10^{-10}\ {\rm eV}^2$.
In particular this means that we can increase
$\langle {\Delta}^0\rangle$ in the above mass matrix, 
eq.~(\ref{mnu}) to give:
\begin{eqnarray}
\langle {\Delta}^0\rangle \simeq 100 \ {\rm eV}\\
\Delta m_{e\mu}^2 \simeq 8 \times 10^{-8} \ {\rm eV}^2\\
m_{{\nu}_e} \simeq m_{{\nu}_{\mu}} \simeq 6 \times 10^{-4} 
\ \mbox{eV}, \ \  m_{{\nu}_{\tau}} \simeq 3 \ {\rm eV}
\end{eqnarray}
Then we have an energy independent solution to the solar neutrino
problem, with a sufficiently heavy ${\nu}_{\tau}$ to provide 
a candidate for hot dark matter. 

We conclude that it is possible to obtain observable neutrino 
oscillations in our model, but only at the expense of introducing 
a new mass scale. It is quite natural to obtain a pair of 
neutrinos with quasi-degenerate masses and a maximum mixing angle 
of $\theta = \pi/4$. Otherwise all mixing angles are expected 
to be small. This is true for a rather general class 
of models which generate the hierarchical structure of the effective 
low energy Majorana neutrino mass matrix by a similar mechanism to 
ours, using new approximately conserved chiral (gauge) charges. 
In particular we do not expect to obtain significant mixing 
between all three neutrinos.

\section{Conclusions}
\label{conclusions}

We have described the development of our proposed system of 
Higgs fields, which generates a realistic quark-lepton mass 
spectrum and breaks the anti-grand unification (AGUT) gauge group 
$SMG^3 \otimes U(1)_f$ down to the SM gauge group SMG.
The most important feature used in this development is that, in our 
AGUT model, the diagonal elements are suppressed to the same 
degree in each of the three charged fermion mass matrices 
$M_U$, $M_D$ and $M_E$. By using this property and consideration of  
possible values for the Abelian gauge quantum numbers, we were led 
to a scheme in which the Planck scale Yukawa couplings for 
the two heaviest generations were fitted in terms of two suppression 
factors $W$ and $T$. A couple of numerical coincidences supported this 
picture and we ended up with the suggested formulae,  
eqs.~(\ref{hmuhs}-\ref{vcbvts}).

These expressions could arise from a general quantum number system, 
sharing a few properties with our AGUT model. The most important 
property is the rule that the same quantum number differences suppress 
the diagonal elements for a given generation in each of the three mass 
matrices. Also the top quark mass is assumed to be unsuppressed by the 
quantum numbers. Finally it has to be checked that the other  
matrix  elements, assumed to be small in the phenomenologically 
suggested Yukawa coupling matrices in terms of $W$ and $T$, 
do not come out to be too large. 
It turned out that our AGUT model could indeed realize the 
general scheme, in terms of two Higgs fields vacuum expectation values 
$\langle W \rangle$ and $\langle T \rangle$ expressed in 
units of the fundamental (Planck) mass scale.

Then, more trivially, the model was extended to include the first 
generation; with all three first generation particles predicted to 
have order of magnitudewise equal masses. 
This was done by introducing 
two more Higgs fields, $\xi$ and $S$, the latter of which however 
gives no suppression, i.e. we took $\langle S \rangle =1$. 
That really means we could have left out the Higgs field $S$ and 
replaced the AGUT gauge group $SMG^3 \otimes U(1)_f$ by the 
subgroup $SMG_{12} \otimes SMG_3 \otimes U(1)$, which survives the 
spontaneous breakdown due to $S$. We managed to naturally include 
the well-known \cite{Fritzsch} order of magnitude relation 
$V_{us} \simeq \sqrt{\frac{m_d}{m_s}}$. It also followed that the 
mass expected for the d quark should be larger, by a factor of 
order 2, than its first order approximation of being degenerate 
with the $u$ quark and the electron at the Planck scale.

We were thereby led to the choice of quantum numbers given in 
eqs.~(\ref{qw}, \ref{qt}, \ref{qxi}, \ref{qs})
for the Higgs fields suppressing the masses, 
and in eq.~(\ref{qws}) for the Weinberg-Salam Higgs field 
of our model. So finally the charged fermion mass matrices 
are given, order of magnitudewise, by the Yukawa matrices 
$H_U$, $H_D$ and $H_E$ of eqs.~(\ref{H_U}-\ref{H_E}).

A fit to the quark-lepton masses and mixing angles was made, 
using a computer program to insert and average over random complex  
numbers of order unity independently multiplying each of 
the matrix elements of $H_U$, $H_D$ and $H_E$. The worst 
deviations from experiment are the strange quark mass, predicted 
to be around 400 MeV or twice its conventional value, and a value of 
the mixing matrix element $V_{cb} \simeq 0.018$ around half its 
experimental value. Since our model only pretends to make order 
of magnitude predictions, even these worst cases should be considered 
as agreement with experiment. We also made a fit using a preliminary 
lattice QCD determination of the light quark masses. Since these 
lattice values are significantly smaller than the conventional values,
and in particular we take the $s$ quark mass to be 85 MeV, 
the second fit is markedly worse; it doubled the average square 
logarithmic deviation between our predictions and the data. 
This result is a rather direct consequence of the simple 
prediction of our model that the order of magnitude of the 
muon mass and the strange quark mass should be the same at the 
Planck scale.

We have also estimated the amount of CP violation in our model. 
Since CP violation arises as a product of rather many of our 
suppression factors, we expect our prediction to be more uncertain 
than that for a typical mass or mixing angle. In fact we predict 
the Jarlskog invariant $J$ measuring CP violation to be a factor of 
3 or 4 below the experimental value. With our expected uncertainty 
in the exponent, this corresponds to approximately one ``standard 
deviation''.

Finally we considered the extension of our model to the neutrino 
sector. The natural see-saw mass scale is the Planck mass in our 
model, for which the only possible observable effect would be 
``just-so'' solar neutrino vacuum oscillations provided the 
electron neutrino $\nu_e$ mixes strongly with $\nu_{\mu}$ or 
$\nu_{\tau}$. The symmetrical Majorana neutrino mass matrix, 
being hierarchical, can naturally give rise to pairs of 
quasi-degenerate neutrino mass eigenstates with maximum mixing. 
In our model, the effective light neutrino Majorana coupling 
matrix eq.~(\ref{hnu}) has this property, but with maximal 
mixing between $\nu_{\mu}$ and $\nu_{\tau}$, while $\nu_e$ 
becomes much lighter than 
and mixes very weakly with the other neutrinos.
So we predict no observable neutrino oscillations, 
unless we modify our model and
introduce another mass scale into the theory. We 
have discussed some ways of introducing a new mass scale 
into our model and given examples with observable 
phenomenological implications. 
These all give neutrino oscillation phenomenology corresponding to
a pair of neutrinos with 
quasi-degenerate masses and a mixing angle $\theta =\frac{\pi}{4}$. 

Apart from the neutrino sector, the AGUT model gives a good
overall order of magnitude fit to the fermion masses and 
mixing angles.

\flushleft{\bf Acknowledgements}

DJS wishes to acknowledge The Royal Society for funding. 
HBN acknowledges funding from INTAS 93-3316, EF contract 
SC1 0340 (TSTS) and Cernf{\o}lge-forskning.
CF acknowledges funding from INTAS 93-3316 and PPARC GR/J21231.

\end{document}